\documentclass[reprint,aps,prd,nofootinbib,a4paper,10pt,twocolumn,superscriptaddress]{revtex4-2}
\usepackage{hyperref} 
\usepackage{changes}
\usepackage{graphicx,subfigure}
\usepackage[section]{placeins}
\usepackage[titletoc]{appendix}
\usepackage{xcolor}
\usepackage{dsfont}
\usepackage{bm}
\usepackage{mathtools} 
\usepackage{amsfonts}
\usepackage{enumerate}

\usepackage{dcolumn}
\usepackage{multirow}
\usepackage{enumitem}

\relax

\hypersetup{
     colorlinks   = true,
     citecolor    = cyan,
     linkcolor = pink
     }
\newcommand\Tstrut{\rule{0pt}{2.6ex}}

\begin{document} 

\title{Model-independent constraints on generalized FLRW consistency relations with bootstrap-based symbolic regression}

\author{S. M. Koksbang} 
\email{koksbang@cp3.sdu.dk}
\affiliation{CP$^3$-Origins, University of Southern Denmark, Campusvej 55, DK-5230 Odense M, Denmark}
\author{A. Heinesen}
\email{asta.heinesen@nbi.ku.dk}
\affiliation{Department of Physics and Astronomy, Queen Mary University of London, UK}
\affiliation{Niels Bohr Institute, Blegdamsvej 17, Copenhagen, 2100, Denmark}

\begin{abstract}
The standard $\Lambda$CDM cosmological model faces increasing tensions between key observations, motivating tests that probe its underlying assumptions. In a companion letter, we present a model‑independent framework that combines derivatives of the angular diameter distance, $d_A(z)$, and the line-of-sight expansion rate, $\mathcal{H}(z)$, to clarify the physical content of FLRW consistency relations and to construct a general‑spacetime estimator of the cosmic density field. Here, we apply these tests to data, introducing a non‑parametric reconstruction method based on symbolic regression combined with bootstrapping to provide data‑driven uncertainty estimates. Using supernova and BAO data, we reconstruct $d_A$, $\mathcal{H}$, and their derivatives, enabling model‑independent evaluation of FLRW relations and recovery of the sky‑averaged density field over $z \in [0.38, \sim 2]$. Current data are too sparse to tightly constrain $\mathcal{H}(z)$, and the reconstructed density is consistent with both Planck and SH0ES $\Lambda$CDM. Reconstructed FLRW consistency tests show mild to moderate deviations from FLRW expectations at the $\sim 2$–$4\sigma$ level, although their significance depends on data selection and reconstruction stability. 
If these indicated deviations from an FLRW geometry are real, it would signify that most of the cosmological solutions considered for solving the cosmological tensions (evolving/interacting dark energy, new types of matter/energy, modified gravity, etc., within the FLRW framework) are ruled out. 
These preliminary indications highlight the importance of future, denser distance and expansion rate measurements, as well as further work toward standardizing uncertainty estimation for symbolic‑regression reconstructions.
\end{abstract}

\maketitle

\section{Introduction}\label{sec1}
The increasing tensions among independent probes of especially late-time cosmology (see e.g. \cite{DiValentino:2021izs}) have motivated a shift from parameter fitting within the $\Lambda$CDM model, to direct tests of its underlying assumptions. Of particular interest are model-independent examinations of the large-scale geometry of spacetime which rely on comparing observationally reconstructed distance measures and the expansion rate with FLRW consistency relations, i.e., relations that must hold in any Friedmann-Lema\^{\i}tre-Robertson-Walker (FLRW) spacetime. Such consistency relations constitute powerful probes of the validity of the FLRW framework and thus the foundation of modern cosmology.
In a companion letter, \cite{PRL}, we highlight a key limitation of these tests. They are derived assuming FLRW geometry. As a result, potential deviations from FLRW expectations cannot readily be interpreted physically. In \cite{PRL}, we overcome this limitation by deriving the FLRW consistency tests for general spacetimes and introducing a new density estimator that also functions as a $\Lambda$CDM test. These relations open the door to a new generation of genuinely diagnostic, model-independent cosmological tests.
\newline\indent
A central obstacle now becomes the handling of observational data. Applying our generalized consistency relations sensibly requires model-independent reconstructions of $d_A$, the line-of-sight expansion rate, $\mathcal{H}$, and their derivatives. The dominant approach for model-independent constraints in cosmology has become Gaussian Process regression (see e.g. \cite{GP1, GP2} for introductions to Gaussian Processes). Gaussian Process reconstructions are powerful, but rely on kernel choices and hyperparameters that can significantly influence the reconstructed functions and, especially, their derivatives \cite{kernel1, kernel2, MAQSOOD2026140456} (see also \cite{GP_model} for considerations on Gaussian Process and model-dependence). These issues become increasingly important as precision cosmology moves toward diagnosing subtle geometric FLRW departures. In this work, we therefore introduce and explore a complementary non-parametric, model-independent method for reconstructing these quantities by combining symbolic regression with bootstrap uncertainty estimation.
\newline\indent
Symbolic regression is a machine learning technique that searches directly for analytical expressions to a dataset, rather than merely fitting parameters of a fixed family of functions (see e.g. \cite{SR2} for an introduction to symbolic regression). Instead of assuming a kernel as in Gaussian Process reconstruction, or assuming a specific fixed family of functions as in traditional regression/fitting approaches, symbolic regression explores large sets of possible algebraic expressions constructed from a set of elementary operations and functions such as addition, division, multiplication, exponential functions, logarithms and trigonometric functions. Candidate expressions are generated, evaluated and iteratively refined according to ranking criteria that are typically based on rewarding both accuracy and simplicity. Since the result of symbolic regression is an explicit functional expression, the reconstructed functions are analytically interpretable and differentiable in closed form. This equation-based structure makes symbolic regression particularly attractive for cosmological consistency tests: First of all, the analytical form facilitates stable derivative estimation which is essential for evaluating most FLRW consistency relations. Second, unlike Gaussian Process reconstruction, symbolic regression does not impose a smoothness prior. Instead, symbolic regression allows the data to determine the functional structure (within the limits of the search space). Third, the analytical forms resulting from symbolic regression could potentially hint at physical structure or systematic effects which would be very difficult to extract from methods such as Gaussian Process reconstruction.
\newline\newline
In the following, we introduce and use our bootstrap-based symbolic regression approach to reconstruct $d_A, \mathcal{H}$ and their derivatives model-independently and asses the corresponding uncertainties. To put these reconstructed quantities to use, we combine them into our newly proposed diagnostic consistency tests. In the first section below, we briefly introduce our diagnostic consistency tests. In section \ref{sec:method} we proceed to describe and test our proposed bootstrap-based symbolic regression approach before we in sections \ref{sec:resultsI} and \ref{ref:resultsII} apply it to supernova and BAO (baryon acoustic oscillations) data. We summarize and conclude in section \ref{sec:conclusion}.

\section{Diagnostic consistency tests} 
In \cite{PRL} we derive expressions for the derivatives of the redshift-derivatives of the angular diameter distance, $d_A'$ and $d_A''$, valid for general spacetimes, requiring only that the relation between the redshift $z$ and the affine parameter $\lambda$ is invertible. With these at hand, we proceed to generalize existing well-known consistency relations so that they are re-expressed in terms of our generalized relations. For the Clarkson-Basset-Lu (CBL) test \cite{cbl} we find that the generalized expression is given by 
\begin{align}\label{eq:Cderived}
    \begin{split}
    \mathcal{C}&= 1 + H^2\left(DD'' - D'^2\right) + HH'DD'\\
    &=1+\frac{d_A^2}{(1+z)^2}\left(-|\hat\sigma|^2-\frac{1}{2}R_{\mu\nu}k^\mu k^\nu + \hat\theta(1+z)\mathcal{H}-\frac{\hat\theta^2}{4}\right)\\& + d_A^2\left( \mathcal{H}\mathcal{H}'(1+z)-\mathcal{H}^2 \right),
    \end{split}
\end{align}
where $\mathcal{H}$ is the line-of-sight expansion rate
\begin{align}
    \mathcal{H} = \frac{1}{3}\theta - e^\mu a_\mu + e^\mu e^\nu\sigma_{\mu\nu}.
\end{align}
Here, $\theta$ is the local expansion rate of the fluid while $a^\mu$ is its acceleration. The shear tensor is given by $\sigma_{\mu\nu}$ and $e^\mu$ is the spatial direction of emission. Similarly to $\theta$ and $\sigma_{\mu\nu}$, we can define an optical expansion rate, $\hat\theta$, and shear tensor, $\hat\sigma_{\mu\nu}$, that describe the deformation of a light beam in the two-dimensional screen space orthogonal to the light ray tangent vector $k^\mu$ and the 4-velocity field of fiducial observers (remember that the optical scalars are observer-independent \cite{sachs}). From $\hat\sigma_{\mu\nu}$ we then define the optical shear scalar as $\hat\sigma^2 = \hat\sigma_{\mu\nu}\hat\sigma^{\mu\nu}$. Lastly, we have denoted the Ricci tensor by $R_{\mu\nu}$.
\newline\indent
As discussed in \cite{asta}, $\mathcal{H}$ is the quantity that we effectively measure as the ``Hubble parameter'' in observations measuring the line-of-sight expansion. It is thus, for instance, explicitly the parameter obtained both from cosmic chronometers \cite{Heinesen:2024gdi} and the longitudinal BAO observations \cite{astabao1, astaboa2}. The Ricci tensor appears in the above as $R_{\mu\nu}$ contracted with the null-tangent vector $k^\mu$, while $\hat\sigma$ and $\hat\theta$ are the optical shear and expansion scalars.
\newline\indent
In \cite{PRL} we further identify the generalized version of the integral of the CBL test as
\begin{align}\label{eq:O}
\begin{split}
    \mathcal{O} &= \frac{\mathcal{H}^2D'^2-1}{D^2} \\&= \frac{\mathcal{H}^2\left( 1+\frac{\hat\theta^2}{4(1+z)\mathcal{H}^2} - \frac{\hat\theta}{(1+z)\mathcal{H}} \right)-1/d_A^2}{(1+z)^2}
    \\&\underset{\mathclap{\text{FLRW}}}{=} \,\,\,\,\,\Omega_{k,0}H_0^2,
\end{split}
\end{align}
where $\Omega_{k,0}$ is the FLRW curvature parameter.
\newline\indent
Lastly, we in \cite{PRL} identify the new useful relation 
\begin{align}\label{eq:density}
\begin{split}
\mathcal{M}:&=\,\,\,\,\,\,\,\,\,\,\,\,-\frac{2\mathcal{H}^2}{3d_A(1+z)}\left(d_A'\left[\frac{2}{1+z}+\frac{\mathcal{H}'}{\mathcal{H}}\right] + d_A''\right)\\&= \frac{2|\hat\sigma|^2+R_{\mu\nu}k^\mu k^\nu}{3(1+z)^5 }\\ &\underset{\mathclap{\text{perfect fluid}}}{=}\,\,\,\,\,\,\,\,\,\,\,\,\frac{2|\hat\sigma|^2}{3(1+z)^5 } + \frac{8\pi G}{3(1+z)^3}(\rho+p)\\ &\underset{\mathclap{\text{$\Lambda$CDM}}}{=} \,\,\,\,\,\,\,\,\,\,\,\,\Omega_{m,0}H_0^2.
\end{split}
\end{align}
The two top lines in the above are interesting in their own right as they provide a way of testing the null-energy condition when shear is negligible. The third line provides a way of constraining the density field through measurements of $d_A$, $\mathcal{H}$ and their derivatives, in the perfect fluid limit and assuming that general relativity holds. The last line provides a new model-independent $\Lambda$CDM consistency relation.
\newline\indent
These generalized FLRW/$\Lambda$CDM consistency relations serve as diagnostic relations since any violation of them can be clearly interpreted. Using the original FLRW consistency relations, a deviation appears only as an abstract anomaly, offering no indication of which underlying assumption fails. In contrast, the generalized expressions decompose the violation into identifiable contributions, thereby revealing the physical origin of any inconsistency rather than merely signaling its presence.
\newline\indent
In the next section, we introduce in detail our proposed bootstrap-based symbolic regression method for constraining $\mathcal{C}, \mathcal{O}$ and $\mathcal{M}$.

\section{Model-independent analyses based on symbolic regression}\label{sec:method}
We wish to apply our new diagnostic consistency relations to supernova and BAO data. To fully utilize the model-independence of the relations, we must constrain $d_A, d_A', d_A'', \mathcal{H}$ and $\mathcal{H}'$ without introducing unnecessary assumptions such as FLRW geometry. As discussed in the introduction, we here propose to use bootstrap-based symbolic regression for this task.
\\\\
Symbolic regression is a machine learning technique that fits data to symbolic expressions without supplying a predefined functional form. Because cosmological data contains noise and some of it is sparse, we should not expect a symbolic regression algorithm to recover an exact analytical representation of the underlying $d_A$ and $\mathcal{H}$ \cite{cp3bench} (see e.g. also \cite{SR1, SR2, SR3, SR4, SR5} for discussions that touch upon the idea of obtaining the underlying ``true'' functional forms using symbolic regression). Instead of aiming for a single expression for each of these, we therefore generate multiple approximations using bootstrap data samples and compute the median and percentiles\footnote{We report median values along with e.g. the 16th and 84th percentiles rather than means and standard deviations because the distributions of the density and CBL test statistics are not expected to be Gaussian.} of their predictions on a redshift grid following the procedure detailed below.
\newline\newline
\subsection{Data}
We wish to reconstruct $d_A, d_A'$ and $d_A''$ from supernova data using the Pantheon+ dataset \cite{pantheon} while $\mathcal{H}$ and $\mathcal{H}'$ will be reconstructed from the BOSS, eBOSS and DESI \cite{DESI:2024mwx} BAO data shown in table \ref{table:BAO} (following \cite{Dinda:2024ktd} for easy comparison with their results) as well as with DESI data release 2 (table IV in \cite{DR2}). As shown in \cite{astabao1, astaboa2}, the radial BAO measurements directly constrain $\mathcal{H}$. Cosmic chronometers data also directly measures $\mathcal{H}$ \cite{Heinesen:2024gdi} and as discussed in \cite{Koksbang:2021qqc, Heinesen:2024gdi}, constraints obtained with cosmic chronometers are entirely independent of FLRW assumptions. However, our initial investigation revealed that cosmic chronometers data still has too large uncertainty to be useful for constraining $\mathcal{H}$ through symbolic regression unless we permit unreasonably large computational resources considering the number of times we need to run the symbolic regression algorithms to obtain the necessary bootstrap samples. We therefore use BAO data, marginalizing over the sound horizon scale $r_s$ as detailed further below. This marginalization may be especially important since mismatches between supernovae and BAO data can occur due to the anchoring of BAO via $r_s$ while supernova data is anchored via SH0ES (see e.g. \cite{anchor1, anchor2}).
\newline\indent
We used Pantheon+ data to constrain
\begin{align}
  f:=\ln(d_A) = \ln(10)/5\cdot (\mu - 25) - 2\ln(1+z) ,
\end{align}
where $\mu$ is the distance modulus. We use the Pantheon+ reported distance moduli directly rather than the raw Pantheon+ data since it was demonstrated in \cite{Dam:2017xqs} that $\alpha, \beta$ and $\Delta_M$ do not vary notably between various test cosmologies. 
We constrained $f$ rather than $d_A$ since $f$, being proportional to $\mu$, has approximately Gaussian errors (assuming errors on the redshift can be neglected). We use redshifts reported in the rest frame of the Cosmic Microwave Background (CMB) but note that this choice should be of little importance for our results since we average across the sky and do not have BAO data below $z= 0.38$ which means that our final constraints of $\mathcal{O}, \mathcal{C}$ and $\mathcal{O}$ are only valid at $z\ge0.38$.

\subsection{Bootstrap-based symbolic regression}
We begin by applying a suite of symbolic regression methods provided through cp3-bench\footnote{https://github.com/CP3-Origins/cp3-bench} \cite{cp3bench} which was designed to implement multiple symbolic regression methods to the same dataset. The algorithms we install through cp3-bench are\footnote{Three other methods are incorporated into cp3-bench, namely  ffx\footnote{https://github.com/ffx-org/ffx/tree/master} \cite{ffx}, gpg\footnote{https://github.com/marcovirgolin/gpg} \cite{gpg} and Operon\footnote{https://github.com/heal-research/pyoperon/} \cite{operon} but the installation of these through cp3-bench is currently broken.}: AIFeynnman\footnote{https://github.com/lacava/AI-Feynman/} \cite{feynman1, feynman2}, DSO\footnote{https://github.com/dso-org/deep-symbolic-optimization} \cite{dso1, dso2}, DSR\footnote{https://github.com/lacava/deep-symbolic-regression} \cite{dsr}, uDSR\footnote{https://github.com/dso-org/deep-symbolic-optimization} \cite{udsr1, udsr2}, GeneticEngine\footnote{https://github.com/alcides/GeneticEngine} \cite{genetic}, gpzgd\footnote{https://github.com/grantdick/gpzgd} \cite{gpz}, ITEA\footnote{https://github.com/folivetti/ITEA/} \cite{itea}, PySR\footnote{https://github.com/MilesCranmer/PySR} \cite{pysr} and QLattice\footnote{https://github.com/abzu-ai/QLattice-clinical-omics} \cite{qlattice}. In general, these algorithms are provided a set of basic operations ($+,-,\cdot, /$) as well as a set of basis functions such as trigonometric, exponential and logarithmic. The algorithms then use different strategies to combine these basis operations and functions and assess how well they approximate the given input data. The strategies are typically based on e.g. genetic algorithms and neural networks, and the assessment of a function is usually based on a combination of two criteria: precision in reproducing the data, and complexity, where more complex functions are penalized. Different algorithms do not in general use the same measures of precision and complexity although mean-square-error is common as a precision measure. Some algorithms installed with cp3-bench are simple for the user to modify in terms of choice of basis functions and hyperparameters, whereas others permit only limited user-friendly adaption customization. When installed with cp3-bench, these are all set in the respective \texttt{procedure.py}-files. For the work presented here, we used the default (none test mode) hyperparameters, except that we adjust the parameters determining how long the algorithms are permitted to run. With our bootstrap procedure we are forced to run each algorithms at least 50 times and thus need to reduce the allowed run time either directly (possible for some algorithms) or indirectly by adjusting hyperparameters such as ``number of epochs".  
The precise hyperparameters used to constrain $\mathcal{H}$ and $d_A$ for the main algorithms we consider are provided in appendix \ref{app:pysr}.
\newline\indent
Once symbolic expressions for $d_A$ and $\mathcal{H}$ are obtained, we can compute predicted values of $d_A, d_A', d_A'', \mathcal{H}$ and $\mathcal{H}'$ on a redshift grid and calculate their median and percentiles. We detail the procedure below, and show results for mock data to verify the robustness of the approach before applying it to real data.
\newline\newline
\begin{table}[!htb]
    \centering
    \begin{tabular}{c c c}
    \hline\hline
    $z_{\rm eff}$ & $c/(\mathcal{H}r_s)$ &  Reference\\
    \hline
    \Tstrut
    $0.510$ & $20.98\pm 0.61$  & \cite{DESI70} \\
    $0.706$ & $20.08\pm 0.60$ & \cite{DESI70}\\
    $0.930$ & $17.88\pm 0.35$ & \cite{DESI:2024mwx}\\
    $1.317$ & $13.82\pm 0.42$ & \cite{DESI71}\\
    $2.330$ & $8.52\pm 0.17$ & \cite{DESI:2024mwx}\\
        \hline
    $0.38$ & $24.981\pm 0.582$ & \cite{BOSS72}\\
    $0.51$ & $22.317\pm 0.482$ & \cite{BOSS72}\\
    $0.698$ & $19.326 \pm 0.469$ & \cite{eBOSS73}\\
    $1.48$ & $13.261 \pm 0.469$ & \cite{eBOSS74}\\
    $2.334$ & $8.99 \pm 0.19$ & \cite{eBOSS75}\\
    \hline
    \end{tabular}
    \caption{BAO data from eBOSS, BOSS and DESI. The top five data points are from DESI DR1 while the bottom five are from BOSS/eBOSS.}
\label{table:BAO}
\end{table}

\begin{figure*}
    \centering
    \includegraphics[width=1\columnwidth]{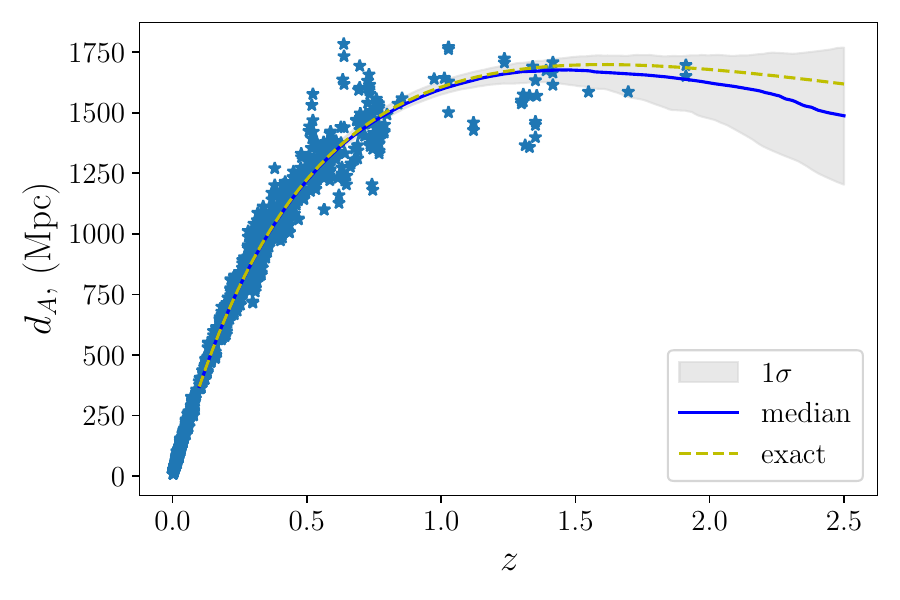}
    \includegraphics[width=1\columnwidth]{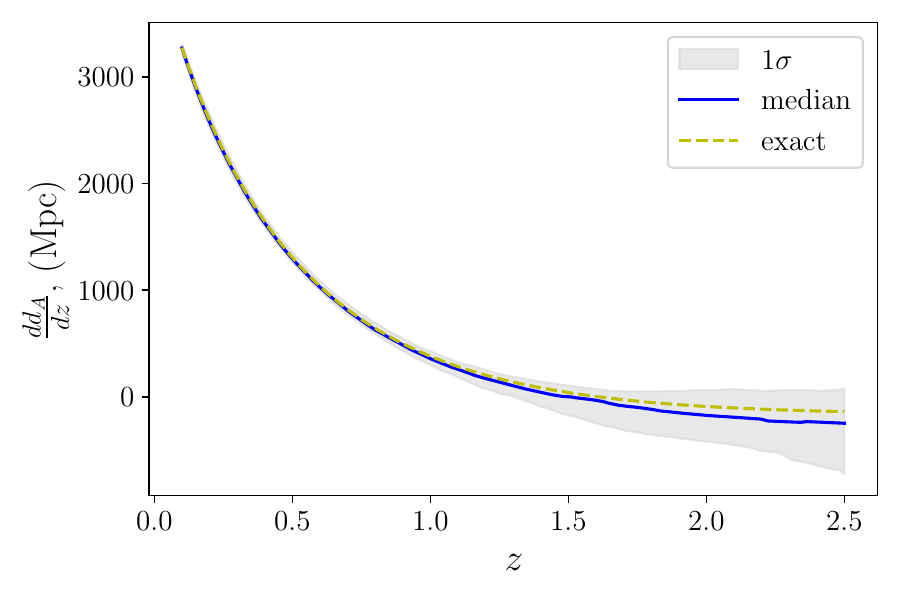}
    \includegraphics[width=1\columnwidth]{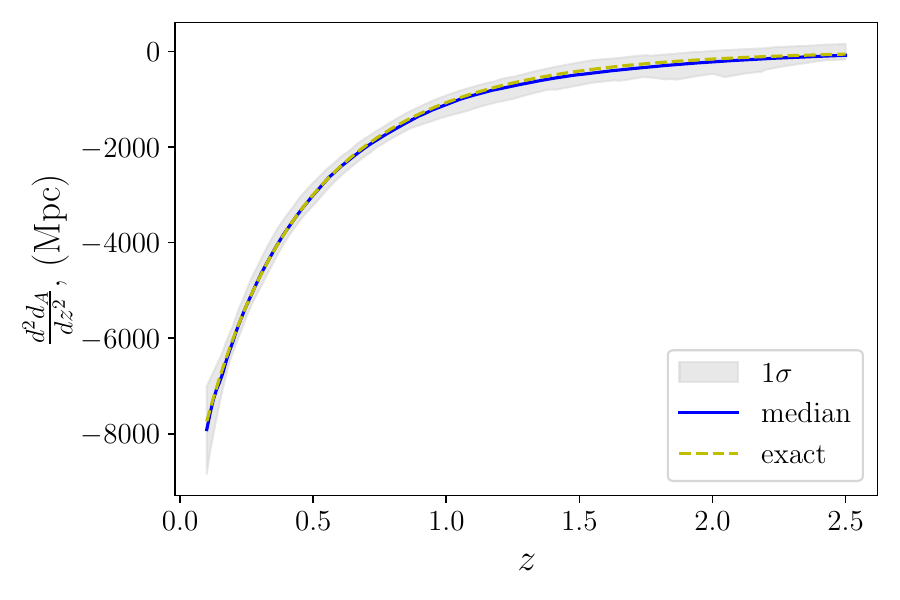}
    \caption{Angular diameter distance and its derivative of the $\Lambda$CDM model with $H_0 = 72$km/s/Mpc. The exact $d_A(z)$ is shown together with a fiducial mock dataset and the median and 16th/84th percentile range (labeled as $1\sigma$) of the symbolic expressions. medians and 16th/84th percentile ranges are also shown for the derivatives.
    }
    \label{fig:mockDA}
\end{figure*}

The first mock measurements we consider are based on a $\Lambda$CDM model with $H_0 = 70$km/s/Mpc and $\Omega_{m,0} = 0.3$. We adopt the same redshift distribution as that of the real data we will consider later\footnote{For both mock and real Pantheon+ data, we remove the first data point, since its large uncertainty causes instability during symbolic regression.} and compute the exact model values of $\ln(d_A)$ and $\mathcal{H}$ (coinciding with the Friedmann Hubble parameter, $H$, in this test case) at each data point. Gaussian noise with mean zero and variance equal to the uncertainty of the corresponding real data point is added to each mock BAO data point. The real data points we use in our analysis are shown in table \ref{table:BAO}. For mock supernova data, we instead use the full covariance matrix of Pantheon+ to generate errors with a multivariate normal distribution.
\newline\indent
For the purpose of making initial, exploratory considerations only, we first generate five bootstrap samples of the mock data by assigning errors in this manner to each sample. These samples are used solely to obtain a preliminary sense of how symbolic regression behaves and to design appropriate selection criteria. They are not used for validation or for drawing any final scientific conclusions. To marginalize over the value of the sound horizon at the baryon drag epoch, $r_s$, we rescale $H$ by $r_{\rm planck}/r_{\rm new}$, where $r_{\rm planck} = 146.995$Mpc and  $r_{\rm new}=r_{\rm planck}+ \delta r$, where $\delta r\in[-2,2]$ is drawn from a uniform distribution for each bootstrap sample. 
\newline\indent
We then perform symbolic regression on each of these five exploratory bootstrap samples using cp3‑bench and manually inspect the resulting symbolic expressions and their derivatives ($d_A', d_A'', \mathcal{H}'$), comparing them to the exact relations used to generate the samples. The purpose of these five samples is exclusively to inform and calibrate the selection criteria. Based on their behavior, we find that accurate reconstructions of $d_A, d_A'$ and $d_A''$ are consistently obtained when we enforce the following criteria for retaining symbolic expressions:\\\\
\textbf{\underline{$d_A$ criteria:}}
\begin{enumerate}[label=\Roman*)]
    \item Only consider expressions with reported (by cp3-bench) mean-square-error smaller than or equal to $0.0110$ (in data units).
    \item Reject all expressions that have clear and strong U or ``check-mark'' shaped first or second derivative.
    \item Remove expressions that have clear oscillations in part or all of the considered redshift interval.
\end{enumerate}
Based on the results from the five bootstrap samples we expect that $d_A$ is reproduced well if the mean-square-error is around 0.00110 or below. In order to reduce the subjectivity of the procedure for selecting $d_A$ expressions detailed below, we therefore enforce a strict cutoff of mean-square-error of 0.001100. This first criterion is typically satisfied by the expressions obtained with AIFeynman, QLattice, ITEA and GeneticEngine. While the remaining algorithms would presumably be able to identify expressions fulfilling this criterion after hyperparameter tuning, we proceed using only these four methods through cp3-bench. The second and third criteria where introduced in order to avoid curves with highly localized features with steep gradients, as the aim is to find the best large-scale cosmological expressions. Note that our criteria are used only to select which expressions to retain among the ``best fit'' expression supplied by each algorithm. Each algorithm has its own set of criteria based e.g. on error and complexity for selecting their ``best fit''.
\newline\newline
We now adopt the above criteria as strict rules for retaining or discarding symbolic expressions when analyzing both real data and a new, substantially larger set of mock data. Having calibrated the selection criteria using the initial five exploratory bootstrap samples, we move on to perform an actual robustness test of the method. To do this, we generate 200 new bootstrap samples based on a $\Lambda$CDM model with $H_0 = 72$km/s/Mpc\footnote{Following best practice, we change the model sightly here since the selection criteria may be over-fitted to the model used to generate the original 5 bootstrap samples.}. According to \cite{bootstrap}, 50-200 samples is sufficient to obtain accurate estimates of the median and uncertainty in most situations. Using these new bootstrap samples, we then follow the procedure described above and summarized here:\\\\
\textbf{\underline{$d_A$ Procedure:}}
\begin{enumerate}
    \item Run cp3-bench for each bootstrap sample and manually inspect the mean-square-error and $d_A', d_A''$ of each resulting symbolic expression. A new random seed is used for each run.
    \item Retain expressions according to $d_A$ criteria I)-III).
    \item Generate new data samples and repeat step 1 and 2 until 200 approved expressions are obtained.
    \item For each expression, calculate the values of $d_A, d_A'$ and $d_A''$ on a redshift grid. Calculate the median and 16th/84th percentile range of the values, sampling over all 200 expressions.
\end{enumerate}
The results of applying this procedure to mock data are shown in Fig. \ref{fig:mockDA}. For all three of $d_A, d_A'$ and $d_A''$, the exact expression for the input $\Lambda$CDM model is within one standard deviation of the median of the symbolic regression results.\\

\begin{figure*}
    \centering
    \includegraphics[width=\columnwidth]{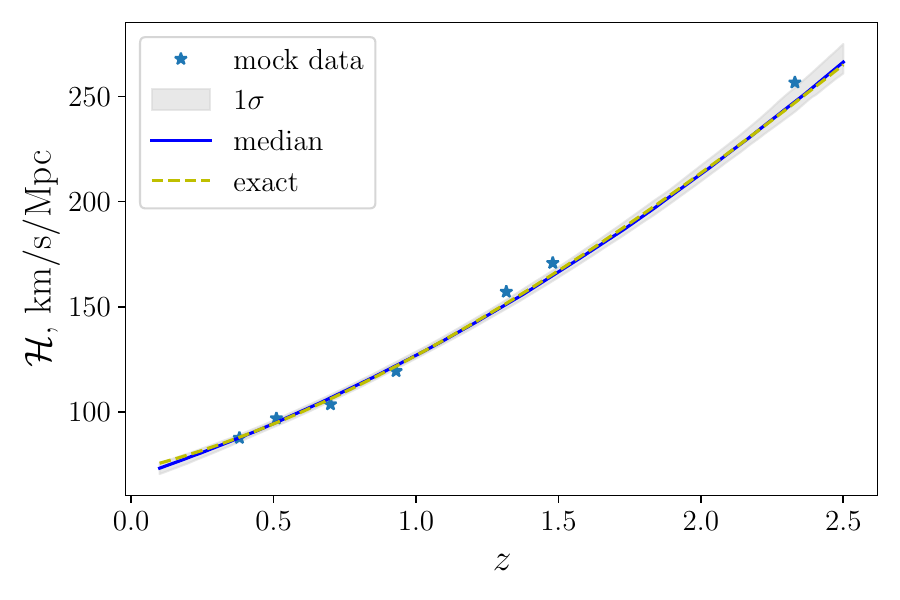}
       \includegraphics[width=\columnwidth]{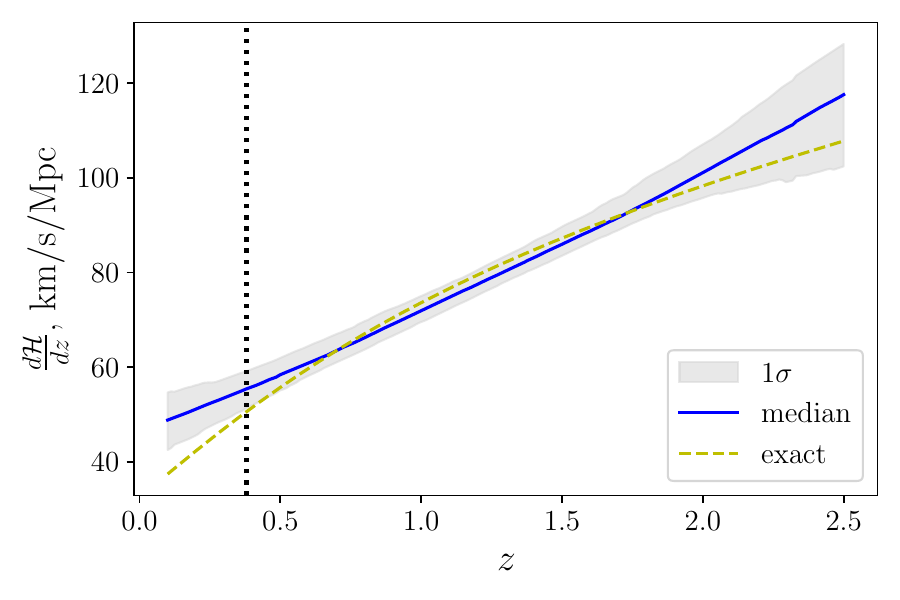}
    \caption{Hubble parameter and its derivative of the $\Lambda$CDM model with $H_0 = 72$km/s/Mpc. The exact $H(z)$ is shown together with a fiducial mock dataset and the median and 16th/84th percentile range of the symbolic expressions. For $H'$,a vertical line is added at the lower redshift range of the data to emphasize that the exact $H$ lies within the 16th/84th percentile range of the median symbolic regression reconstruction in the entire redshift range of the mock data.
    }
    \label{fig:mockH}
\end{figure*}
\begin{figure*}
    \centering
    \includegraphics[width=2\columnwidth]{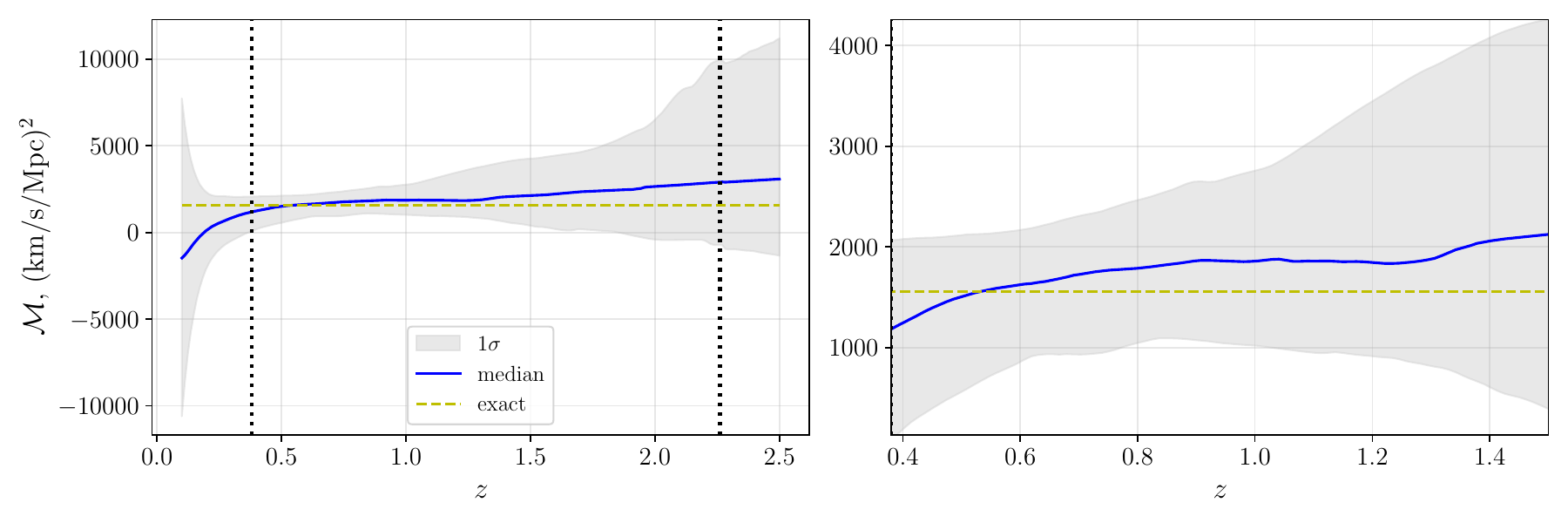}
    \caption{The density test, $\mathcal{M}$, (eq. \ref{eq:density}) based on the mock data for a $\Lambda$CDM model with $H_0 = 72$km/s/Mpc. The exact value is shown together with the median and 16th/84th percentile range (labeled $1\sigma$) obtained with the symbolic expressions. Dashed lines are added to show the redshift range covered by both datasets. The figure to the right show a zoom of the left figure.
    }
    \label{fig:mocktest}
\end{figure*}
\begin{figure*}
    \centering
    \includegraphics[width=2\columnwidth]{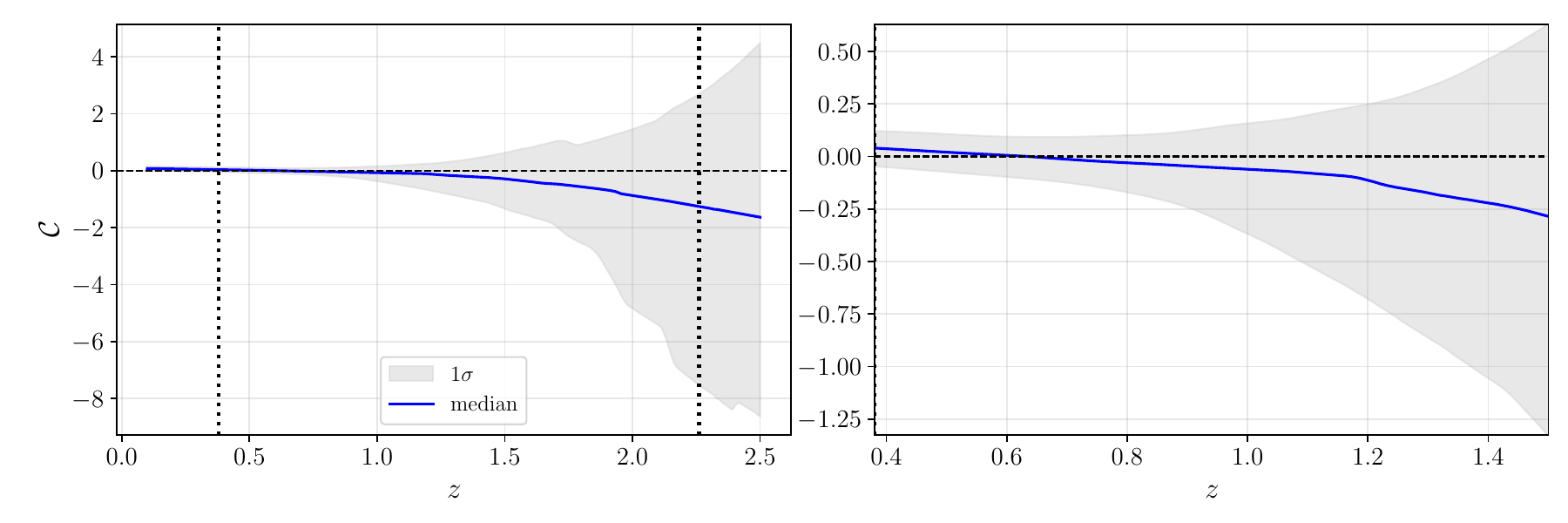}
    \caption{The CBL-test, $\mathcal{C}$, (eq. \ref{eq:Cderived}) based on the mock data for a $\Lambda$CDM model with $H_0 = 72$km/s/Mpc. The exact value is shown together with the median and 16th/84th percentile range (labeled $1\sigma$) obtained with the symbolic expressions. Dashed lines are added to show the redshift range covered by both datasets. The figure to the right shows a zoom of the left figure.
    }
    \label{fig:mockCBLtest}
\end{figure*}
\begin{figure*}
    \centering
    \includegraphics[width=2\columnwidth]{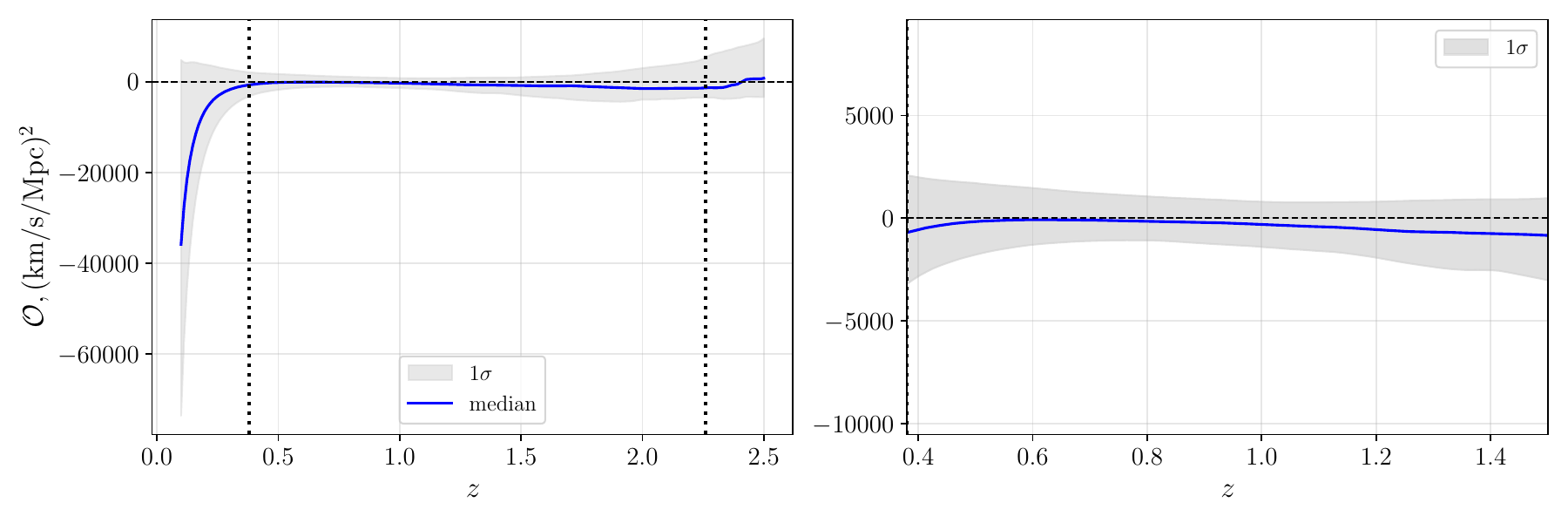}
    \caption{The integrate CBL-test, $\mathcal{O}$, (eq. \ref{eq:O}) based on the mock data for a $\Lambda$CDM model with $H_0 = 72$km/s/Mpc. The exact value is shown together with the median and 16th/84th percentile range (labeled $1\sigma$) obtained with the symbolic expressions. Dashed lines are added to show the redshift range covered by both datasets. The figure to the right shows a zoom of the left figure.}
    \label{fig:mockO}
\end{figure*}

We apply a similar procedure to the BAO data used for constraining $H$. We initially considered the 10 data points in table \ref{table:BAO}. However, six of the data points occur in closely spaced redshift pairs with significantly different values of $H(z)$. Including both data points in such a redshift pair causes severe overfitting, which for two data points with approximately the same $z$ value but widely different $H$ values means that the resulting functions include divergences near these redshifts to permit the functions to trace both data points. We therefore remove one data point in each pair, always retaining the non-DESI data points when producing mock data. When considering real data in the next section, we show results obtained with two versions of the dataset, retaining the DESI and BOSS/eBOSS data points, respectively. We emphasize that this split of data is not ``cherry picking'' of data, but rather a necessity enforced by the nature of symbolic regression algorithms and our wish to avoid diverging expressions.
\newline\indent
From the first, exploratory five bootstrap samples of the mock dataset for $H(z)$ (with $H_0 = 70$km/s/Mpc) we find that most methods installed with cp3-bench struggle to fit the sparse BAO data. Rather than manually tuning the hyperparameters of each method, we proceed with PySR which is easily tunable and which showed the most promise during initial tests. Since we now rely on only a single method, we install it outside cp3-bench and work with PySR directly. After tuning the PySR hyperparameters slightly (see appendix \ref{app:pysr}), we run it on a few bootstrap samples and manually inspect its resulting ``hall-of-fame'' of symbolic expressions, comparing them and their derivatives with the true $\Lambda$CDM $\mathcal{H}=H$ and $\mathcal{H}'=H'$. Based on this inspection, we introduce the following procedure:\\\\
\textbf{\underline{$\mathcal{H}$ procedure:}}
\begin{enumerate}[label=\roman*)]
    \item Run PySR on each bootstrap sample dataset to obtain a hall-of-fame.
    \item From the hall-of-fame, choose expressions with PySR's reported ``Loss'' less than 3.5 and that are well-behaved and have well-behaved first derivatives on the entire training interval and where neither the expression itself nor its derivative are radically U- or check mark-shaped.
    \item Repeat step i) and ii) until 200 different symbolic expressions are obtained (each run yields 0-3 expressions fulfilling the criteria for retention. We therefore end up sampling well above 50 different values of $r_{\rm new}$ and thus expressions based on $>50$ different bootstrap samples).
\end{enumerate}
The criteria in step ii) are necessary since PySR tends to severely overfit the data to obtain very small Loss. Many of these over-fitted expressions and/or their derivatives have wildly unrealistic peaks and/or are undefined on parts of the relevant redshift interval. The Loss criterion is slightly changed for BAO data where the DESI measurements are retained. The error and redshifts slightly change in this case, and a test showed that we should in this case retain expressions up to a loss of 60 while keeping the rest of the criteria and procedure as above. Note however, that the exact value of the Loss threshold is insignificant for the resulting family of expressions. The effect of introducing the threshold together with the other criteria is that we roughly end up choosing functions in the ``middle'' part of Pareto-front, i.e. those with medium values of complexity and of Loss. Together, the criteria ensure that we avoid A) the lowest complexity and highest Loss functions in the pareto-front, which are functions that are constant or linear, and B) the highest complexity and lowest Loss functions which, either themselves or their first derivatives, are non-defined/divergent or wildly oscillating within their training region. In a later section, we demonstrate that choosing other selection criteria has only small impact on the final median and percentiles.
\newline\indent
From the 200 expressions that the above procedure yields, we calculate $\mathcal{H}$ and $\mathcal{H}'$ on a grid and calculate the corresponding median and 16th/84th percentile range. The results are shown in Fig. \ref{fig:mockH}. As seen, the symbolic expressions obtained for $\mathcal{H}$ are not nearly as accurate as those obtained for $d_A$. This is in line with the BAO dataset being very sparse compared to the supernova dataset and with the fact that we chose to be model-agnostic (to the extent possible when using BAO data) and marginalize over $r_s$ rather than fix it to the Planck value. Nonetheless, the reconstructions agree with the exact $\mathcal{H}$ and $\mathcal{H}'$ within the 16th/84th percentile ranges in nearly the entire redshift interval of the data except very close to the lower boundary, where $\mathcal{H}'$ shifts slightly below.
\\\\
Finally, as we have now reconstructed $d_A, d_A', d_A'', \mathcal{H}$ and $\mathcal{H}'$, we can calculate the corresponding density test, $\mathcal{M}$, of Eq. \ref{eq:density}. For the reconstruction, we calculate $\mathcal{M}$ for all expressions where we make all $4\cdot 10^4$ combinations of the symbolic expression for $d_A$ and $\mathcal{H}$\footnote{As an extra robustness test, we have verified that the results are quantitatively unchanged if we instead randomly pair expressions for $d_A$ and $\mathcal{H}$ to obtain exactly 200 pairs. The only difference between the results based on 200 versus 40,000 pairs is that the median and percentiles are \emph{slightly} smoother in the latter case.}. For each of these pairs, we calculate the value of the density test on a redshift grid. The median and 16th/84th percentile range obtained on the grid for these 40 thousand combinations are shown in Fig. \ref{fig:mocktest}. As seen, the median value is not quite constant, but the exact value is easily contained within the 16th/84th percentile range throughout the entire redshift region covered by both mock datasets. The robustness test with mock data thus indicates that the true $\mathcal{M}$ should fall within one standard deviation of the mean result, but the overall shapes of the median and percentiles do not reflect the shape of the actual density field.
\newline\indent
We also combine the reconstructed $d_A, d_A', d_A'', \mathcal{H}$ and $\mathcal{H}'$ to estimate the generalized  CBL test, $\mathcal{C}$. Fig. \ref{fig:mockCBLtest} shows the median and 16th/84th percentile calculated as for $\mathcal{M}$. As seen, the correct model result, $\mathcal{C} = 0$, is contained well within the 16th/84th percentile range around the median in the entire redshift interval of the mock data.
\newline\indent
Lastly, we apply the procedure to constrain the integrated CBL test, $\mathcal{O}$. The results are shown in Fig. \ref{fig:mockO}. Again, we see that the correct value for the input model, $\mathcal{O} = 0$, is contained within the uncertainty band corresponding to $1\sigma$ (i.e. the 16th/84th percentile range).
\newline\newline
Overall, the FLRW test gives us good reason to expect the true functional relations for $d_A, \mathcal{H}$, their derivatives as well as $\mathcal{C}, \mathcal{O}$ and $\mathcal{M}$ will be correctly constrained by bootstrap-based symbolic regression. We thus proceed by applying the method to real data.

\section{Diagnostic consistency relations: Numerical constraints with real data I}\label{sec:resultsI}
We now apply the procedures described in the previous section to real data. We do this in two rounds. In this section, we first consider BAO data from BOSS, eBOSS and DESI data release 1. In the next section, we will consider the newest BAO data from DESI.
\newline\indent
When applying the bootstrap-based symbolic regression method to data in this section, we stick strictly to the selection criteria detailed above and only remove expressions that clearly do not meet the criteria. Both when considering real data and mock data for the robustness test, we do not compare with any model or data values when selecting symbolic expressions. We analyze the Pantheon+ supernovae data (including SH0ES)\footnote{https://github.com/PantheonPlusSH0ES/DataRelease} \cite{pantheon} to obtain model-independent estimates of $d_A, d_A'$ and $d_A''$, and BAO data from DESI \cite{DESI:2024mwx}, BOSS \cite{BOSS72} and eBOSS \cite{eBOSS73} to obtain $\mathcal{H}$ and $\mathcal{H}'$. As discussed in the previous section, we make two BAO data combinations: one where parts of the DESI measurements are removed, and one where parts of eBOSS/BOSS data points are removed due to overlap in redshifts of the DESI/eBOSS/BOSS data points.
\newline\newline

\begin{figure*}
    \centering
    \includegraphics[width=1\columnwidth]{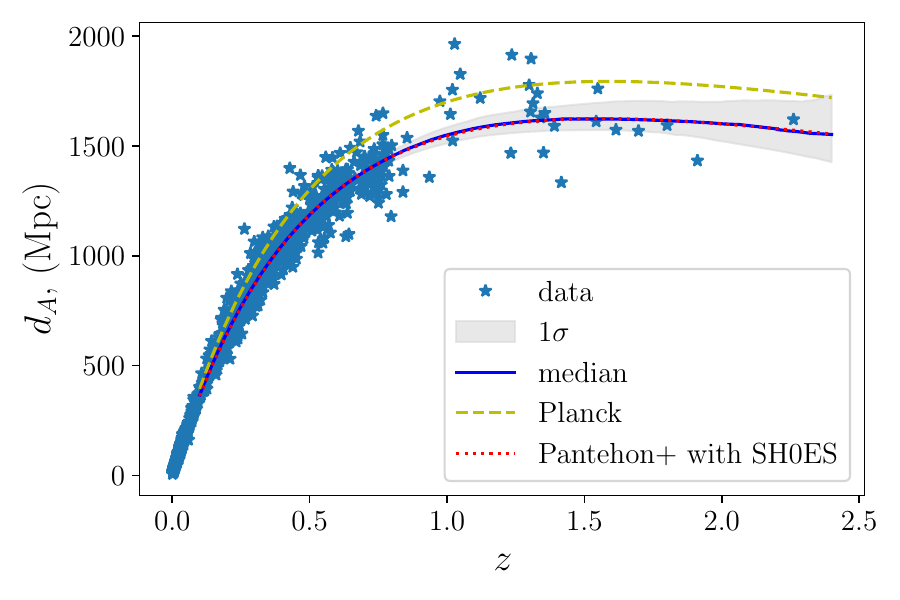}
        \includegraphics[width=1\columnwidth]{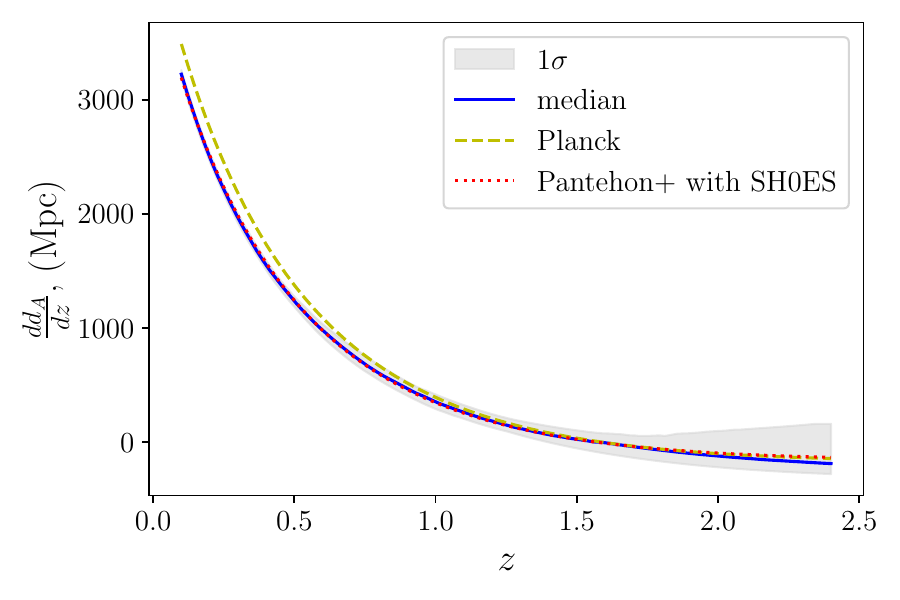}
            \includegraphics[width=1\columnwidth]{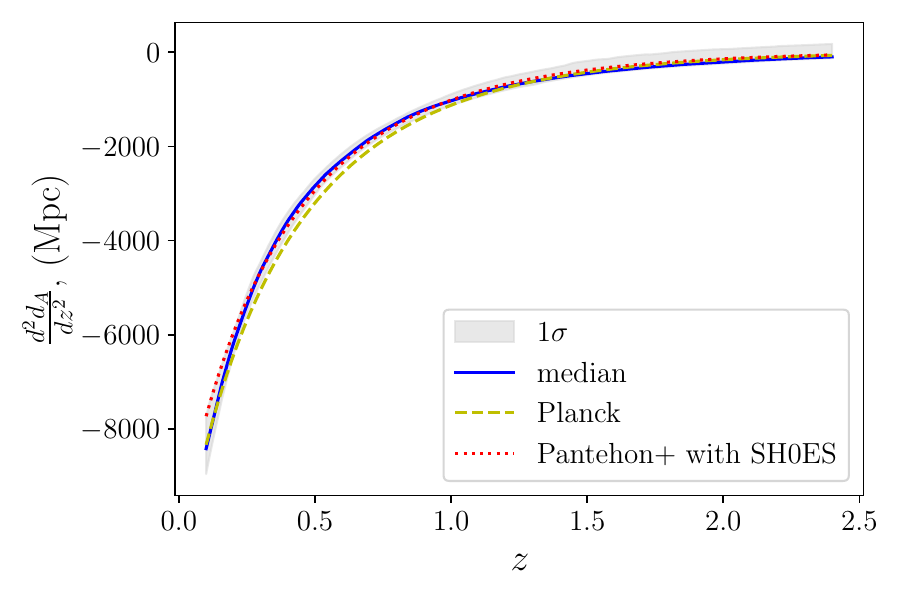}
    \caption{Median and standard deviation for the angular diameter distance and its first and second derivatives for Pantheon+SH0ES data. The reconstructions are shown together with $\Lambda$CDM results using Planck and SHOES values. The used Pantheon+ data is shown in the plot with $d_A$.
    }
    \label{fig:data_DA}
\end{figure*}
\begin{figure*}
    \centering
    \includegraphics[width=\columnwidth]{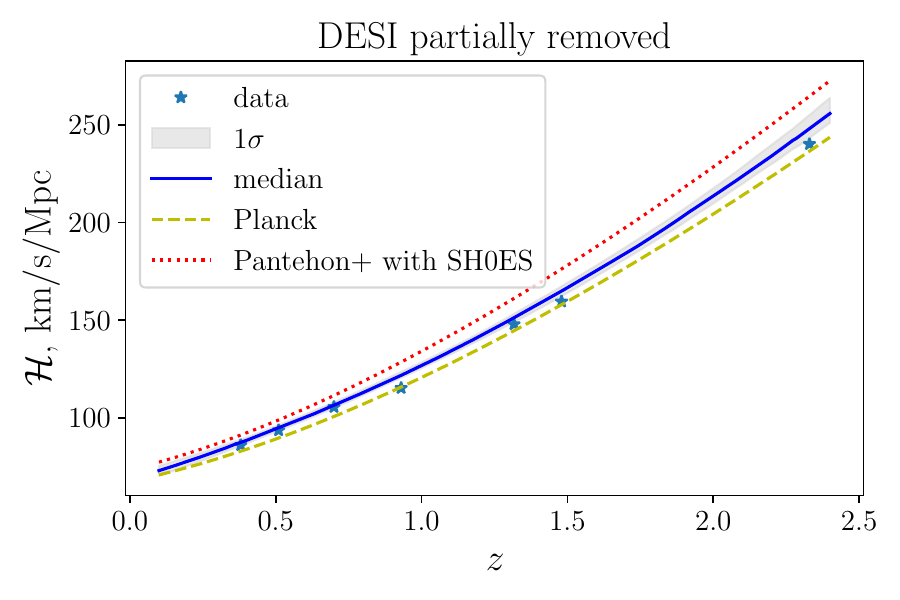}
    \includegraphics[width=\columnwidth]{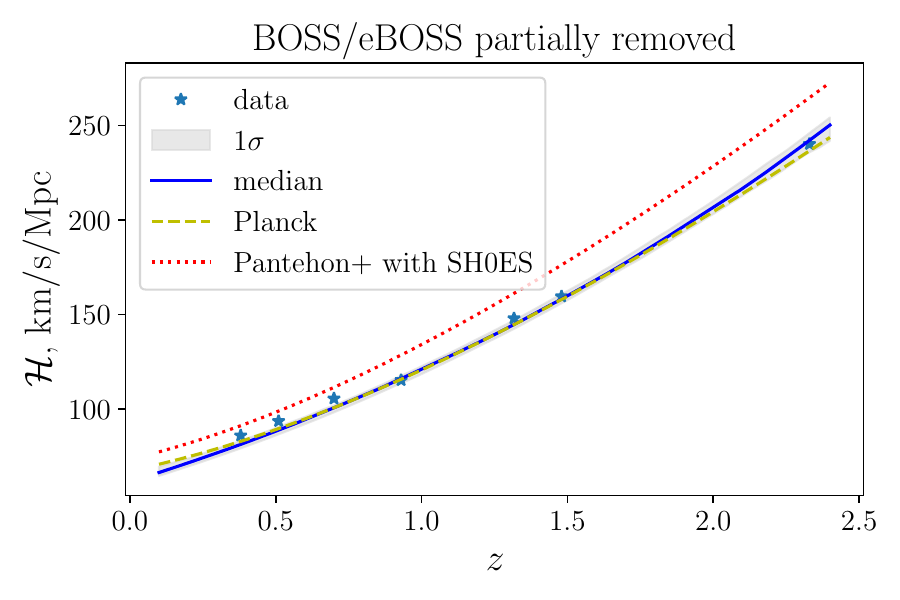}\\
    \includegraphics[width=\columnwidth]{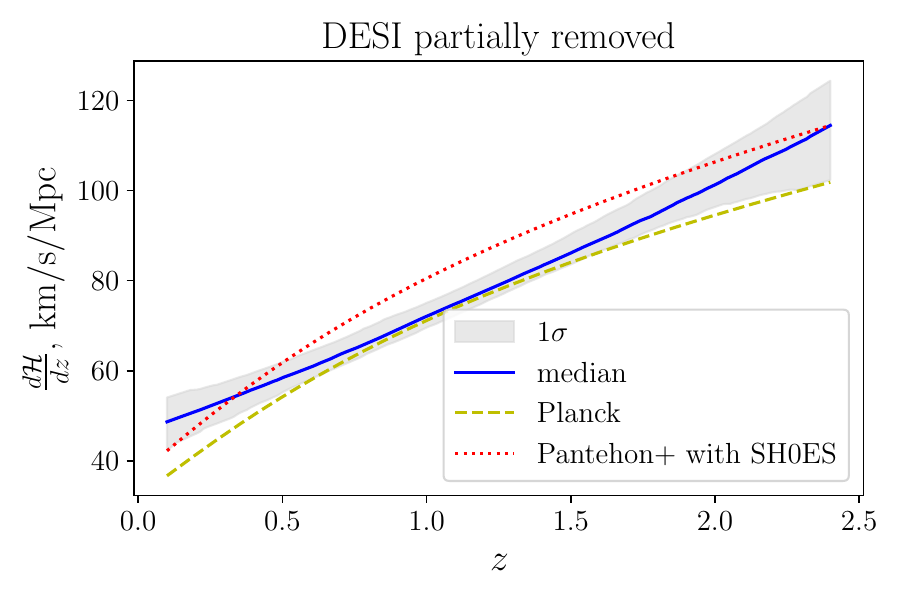}
    \includegraphics[width=\columnwidth]{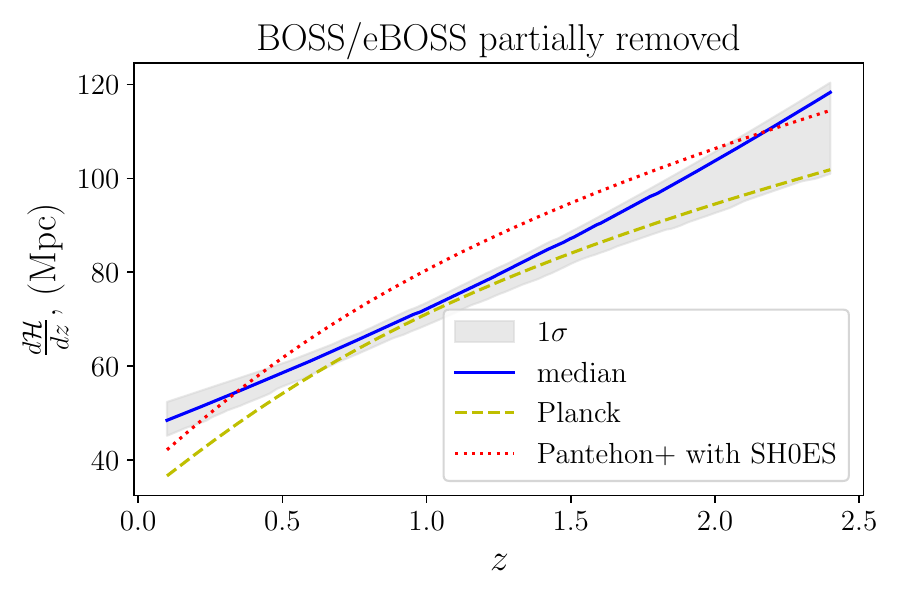}\\
        
    \caption{Reconstructed $\mathcal{H}$ and its first derivative compared with the best-fit Planck value and best-fit value from Pantheon+ combined with SH0ES. The reconstruction is shown as the median and 16th/84th percentiles (labeled $1\sigma$). The reconstruction is based on both versions of the BAO dataset. Data is shown assuming $r_s = 146.995$Mpc (Planck value).
    }
    \label{fig:data_H}
\end{figure*}
\begin{figure*}
    \centering
    \includegraphics[width=2\columnwidth]{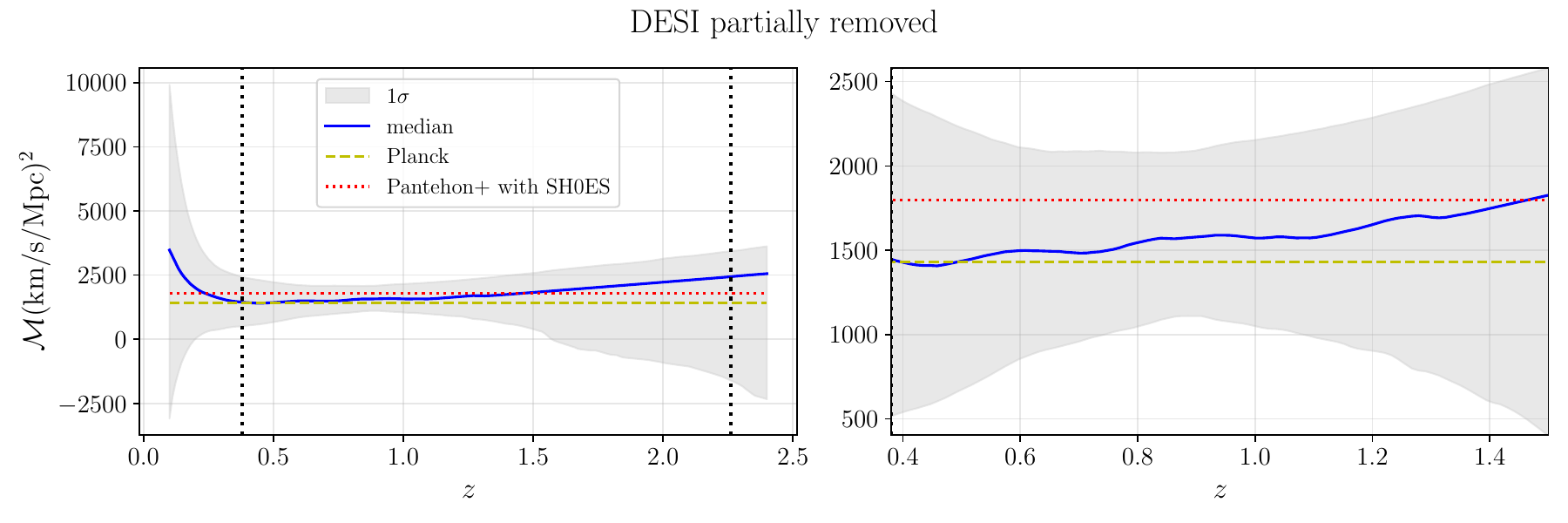}
    \caption{The density test $\mathcal{M}$ (eq. \ref{eq:density}) based on supernovae and BAO data, where three DESI data points have been removed due to redshift overlap with eBOSS/BOSS data points. The best-fit Planck value and best-fit value based on Pantheon+ and SH0ES are shown together with the median and 16th/84th percentiles (labeled $1\sigma$) obtained with the symbolic expressions. Dashed lines are added to show the redshift interval where both datasets contain data. The right-hand plot shows a close-up.
    }
    \label{fig:BOSStest}
\end{figure*}
\begin{figure*}
    \centering
    \includegraphics[width=2\columnwidth]{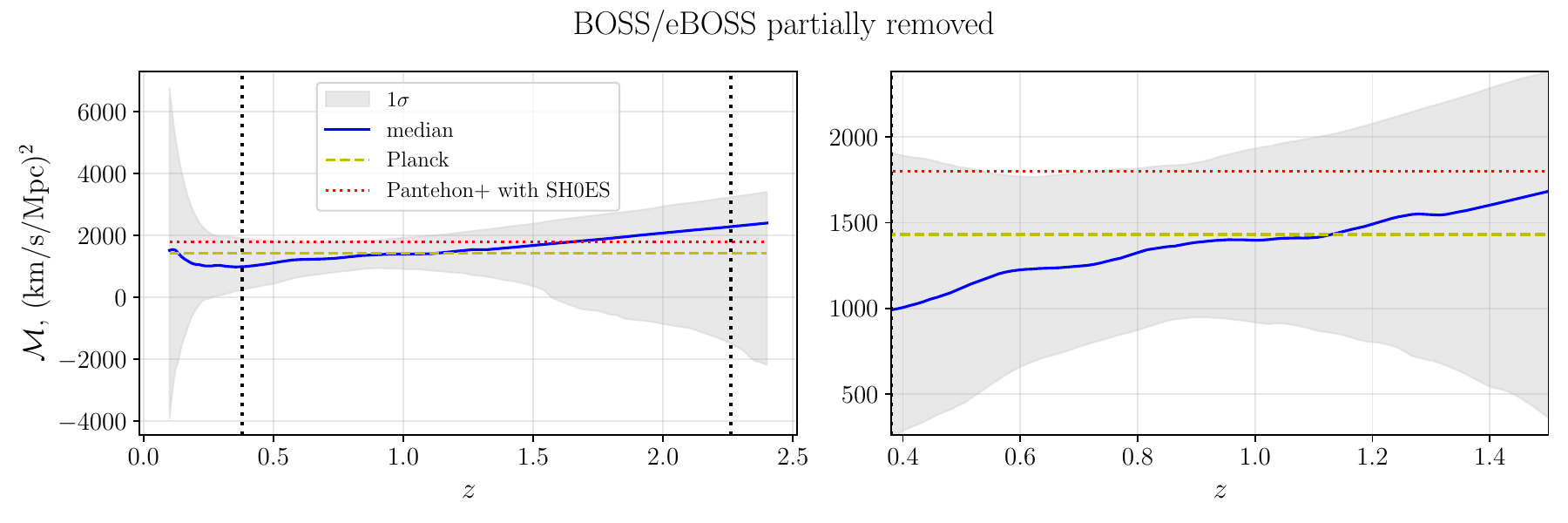}
    \caption{The density test $\mathcal{M}$ (eq. \ref{eq:density}) based on supernovae and BAO data, where three (e)BOSS data points have been removed due to redshift overlap with DESI data points. The best-fit Planck value and best-fit value based on Pantheon+ and SH0ES are shown together with the median and 16th/84th percentiles (labeled $1\sigma$) obtained with the symbolic expressions. Dashed lines are added to show the redshift interval where both datasets contain data. The right-hand plot shows a close-up.
    }
    \label{fig:DESItest}
\end{figure*}
\begin{figure*}
    \centering
    \includegraphics[width=2\columnwidth]{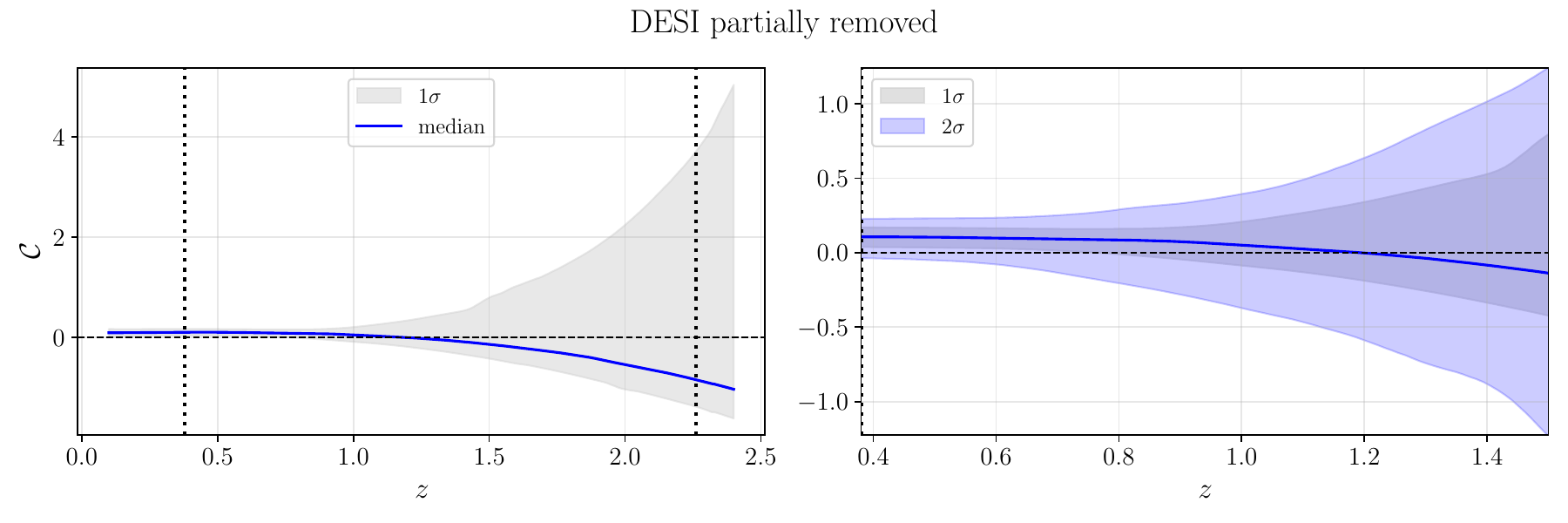}
    \caption{The CBL test $\mathcal{C}$ (eq. \ref{eq:Cderived}) based on supernovae and BAO data, where three DESI data points have been removed due to redshift overlap with DESI data points. We show the median and 16th/84th percentiles (labeled $1\sigma$) obtained with the 200 identified symbolic expressions. Dashed lines are added to show the redshift interval where both datasets contain data. The right-hand plot shows a close-up.
    }
    \label{fig:BOSS_cbl}
\end{figure*}
\begin{figure*}
    \centering
    \includegraphics[width=2\columnwidth]{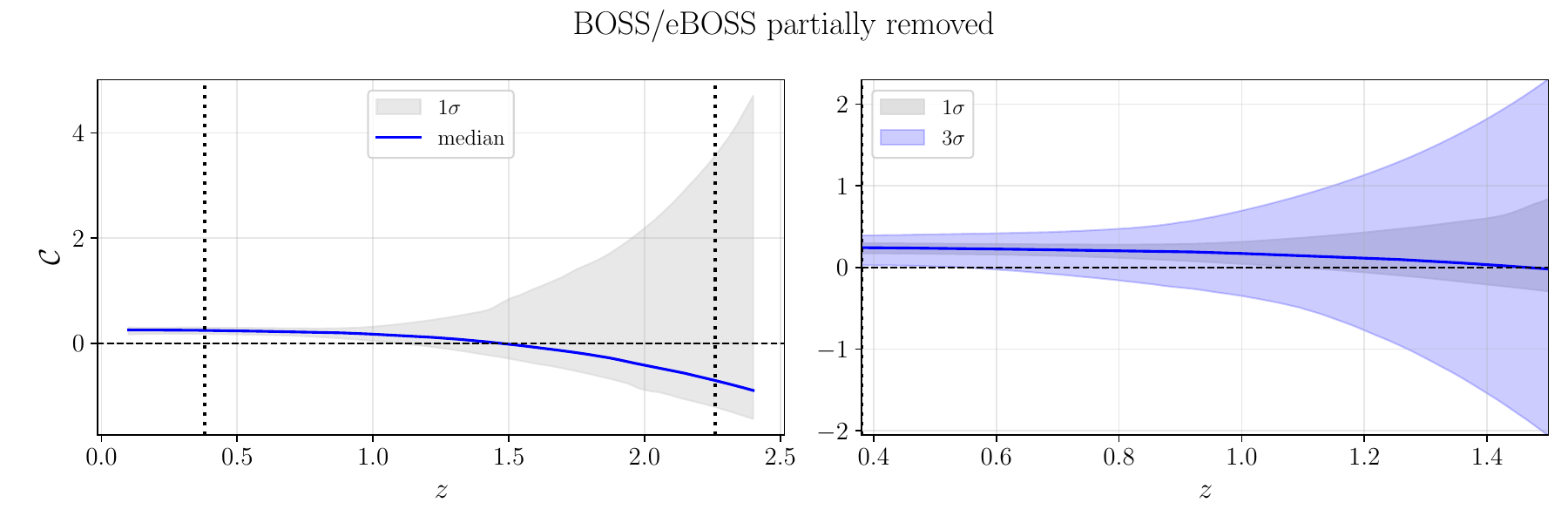}
    \caption{The CBL test $\mathcal{C}$ (eq. \ref{eq:Cderived}) based on supernovae and BAO data, where three (e)BOSS data points have been removed due to redshift overlap with DESI data points. We show the median together with 16th/84th percentiles (labeled $1\sigma$) obtained with the 200 identified symbolic expressions. Dashed lines are added to show the redshift interval where both datasets contain data. The right-hand plot shows a close-up which includes a 1.35th/99.86th percentiles range corresponding to $3\sigma$ for Gaussian distributed values.
    }
    \label{fig:DESI_cbl}
\end{figure*}

The results from applying the bootstrap-based symbolic regression to supernova data is shown in Fig. \ref{fig:data_DA}. As seen, the reconstructed $d_A$ and its derivatives fit very well with the $\Lambda$CDM prediction using Pantheon+SH0ES cosmological parameters while the prediction using the $\Lambda$CDM model with Planck values deviates from the prediction (and the data) very clearly.
\newline\indent
The reconstructions obtained from applying our bootstrapped symbolic reconstruction to BAO data are shown in Fig. \ref{fig:data_H}.
\newline\newline
Once we have constrained $d_A, d_A', d_A'', \mathcal{H}$ and $\mathcal{H}'$, we combine them according to the density-test, $\mathcal{M}$, introduced in Eq. \ref{eq:density}. The result is shown in Fig. \ref{fig:BOSStest} and  \ref{fig:DESItest} where we compare with the Planck estimate $\Omega_{m,0}H_0^2 = 1431.4\rm (km/s/Mpc)^2$ (table 1 of \cite{planck}) and the estimate obtained by combining Pantheon+ and SH0ES, $\Omega_{m,0}H_0^2= 1799.4 \rm (km/s/Mpc)^2$ \cite{Brout:2022vxf}. The figures include vertical dashed lines to delineate the redshift interval where the supernova and BAO datasets overlap. At lower and higher redshifts, the results cannot be trusted since symbolic regression does not in general extrapolate well outside the training interval. As seen, both the Planck and Pantheon+ with SH0ES \cite{Brout:2022vxf} estimates of $\Omega_{m,0}H_0^2$ are within the 16th/84th percentiles ($\sim 1\sigma$ if distributions were Gaussian) of the median symbolic regression estimate in nearly the entire studied redshift interval. The exception is the Pantheon+ and SH0ES estimate which falls outside the 16th/84th percentile range around $z = 0.6$ for the data combination where BOSS/eBOSS data has been partially removed. For both considered data combinations, the Planck value fits well within the 16th/86th percentile range of the model-independent reconstruction of the density test. The Pantheon+ (with SH0ES) prediction, on the other hand, only lies well within the interval when DESI data is partially removed. This is in good agreement with the reconstructed $\mathcal{H}$ which is shown in Fig. \ref{fig:data_H}. We also note that when BOSS/eBOSS data is partially removed, the Planck prediction lies roughly within the 16th/84th percentile of the median symbolic regression result. In all other cases shown, the $\Lambda$CDM predictions lie clearly outside the 16th/84th percentile range of the symbolic regression median prediction of $\mathcal{H}$.
\newline\newline
With reconstructions of $d_A, d_A', d_A'', \mathcal{H}$ and $\mathcal{H}'$ at hand we can also constrain $\mathcal{C}$ which we show in Fig. \ref{fig:BOSS_cbl} and \ref{fig:DESI_cbl}. Interestingly, the FLRW prediction $\mathcal{C} = 0$ is not contained within the 16th/84th percentile range of the median in the larger part of the redshift interval covered by both BAO datasets.
However, when DESI measurements are partially removed, the reconstruction is shifted significantly closer to 0, and 0 is contained in the $2\sigma$ (2.275/97.725 percentiles) interval around the median. When DESI data dominate the BAO data, the disagreement with the FLRW prediction is between 3$\sigma$ and 4$\sigma$. This result differs from that of \cite{list} where the CBL test multiplied by $H^3$ (their $W_{\rm K2}$) was included among several FLRW consistency tests and found to agree with $\mathcal{C} = 0$ within $1\sigma$ (see their Fig. 10). There are several differences between our analysis and that in \cite{list} which can explain this discrepancy. First, \cite{list} uses a slightly different BAO dataset, fixes $r_s$ to its best-fit Planck value, and constrains $W_{\rm K2}$ solely with BAO data. In addition, their treatment of $H'$ and $H''$ (their Eq. 4.4 and 4.5) relies on FLRW geometry, which could introduce bias. Similarly, their Gaussian Process reconstruction assumes FLRW-like relations between $H$ and distance measures (their Eq. 4.1). Finally, Gaussian Process regression itself may favor FLRW-like behavior through kernel choices and hyperparameters that enforce smoothness and monotonicity (see e.g. \cite{kernel1, kernel2, GP_model, MAQSOOD2026140456} for discussions of kernel choices). Identifying the exact reason for the discrepancy between our results and those obtained in \cite{list} is beyond the scope of the current work. Nonetheless, based on the above considerations and exploring constraints obtained with Gaussian Process methods in general, we suspect that the main reason for the discrepancy is the difficulty of using Gaussian Processes to reconstruct derivatives. In \cite{list}, the derivative $d_A'$ can be somewhat well constrained because the Gaussian Process reconstruction is informed that $d_A'\propto 1/H$. Beyond helping to constrain $d_A'$, this explicitly means that FLRW geometry was assumed in the reconstruction which naturally drives the results of \cite{list} to be consistent with FLRW expectations. While the assumption can alleviate the difficulty in reconstructing $d_A'$, it does not help with reconstructing $d_A''$ or $H'$ and poor constraints on these may increase the uncertainty bands in \cite{list}. If a Gaussian Process reconstruction struggles to reconstruct derivatives, this instability manifests as broad uncertainty bands. For symbolic regression it would instead manifest as the appearance of many oscillatory solutions. Although we do find some of these in our Hall-of-fame's, they do not dominate, presumably because symbolic regression algorithms tend to discard highly oscillatory or pathological expressions through built-in complexity penalties. This may explain why our uncertainty bands are narrower than those in \cite{list}.
\newline\indent
The above discussion also highlights the important point that symbolic regression reconstructions, just as Gaussian Process reconstructions, depend on hyperparameter choices. For instance, for some algorithms we can explicitly control the appearance of oscillatory behavior by including/removing sine and cosine in/from the set of basis functions provided to an algorithm. While we indeed did this when reconstructing $\mathcal{H}$, this was not an option for all algorithms we used through cp3-bench to reconstruct $d_A$. Similarly, we can to some extent control the appearance of pathological functions. Again, this is more easily done for PySR than the algorithms we used for reconstructing $d_A$. For instance, when reconstructing $\mathcal{H}$, we did not permit division; when division was permitted, test runs showed that most resulting symbolic expressions included division by zero in the considered redshift interval. For some algorithms, we can further explicitly constrain the permitted nestedness of basis functions which can be used to reduce the risk of e.g. taking the square root of a negative number.
\newline\indent
Overall, each choice of hyperparameters constrains the freedom of the algorithms and thereby places boundaries on the space of admissible symbolic expressions. In this sense, symbolic regression inevitably introduces a form of modeling, albeit not a cosmological model. Different hyperparameter choices can therefore be expected to lead to different outcomes in our bootstrap-based symbolic regression analyses. However, by performing the reconstruction using multiple algorithms, varying hyperparameters, and applying different selection criteria, we aim to reduce the impact of any single ``modeling choice''. Moreover, the hyperparameter choices made in symbolic regression are largely explicit and transparent, in contrast to those underlying Gaussian Process reconstructions, where the modeling assumptions are often less directly accessible  (see e.g. \cite{GP_model} for a discussion of this point).

\begin{figure*}
    \centering
    \includegraphics[width=2\columnwidth]{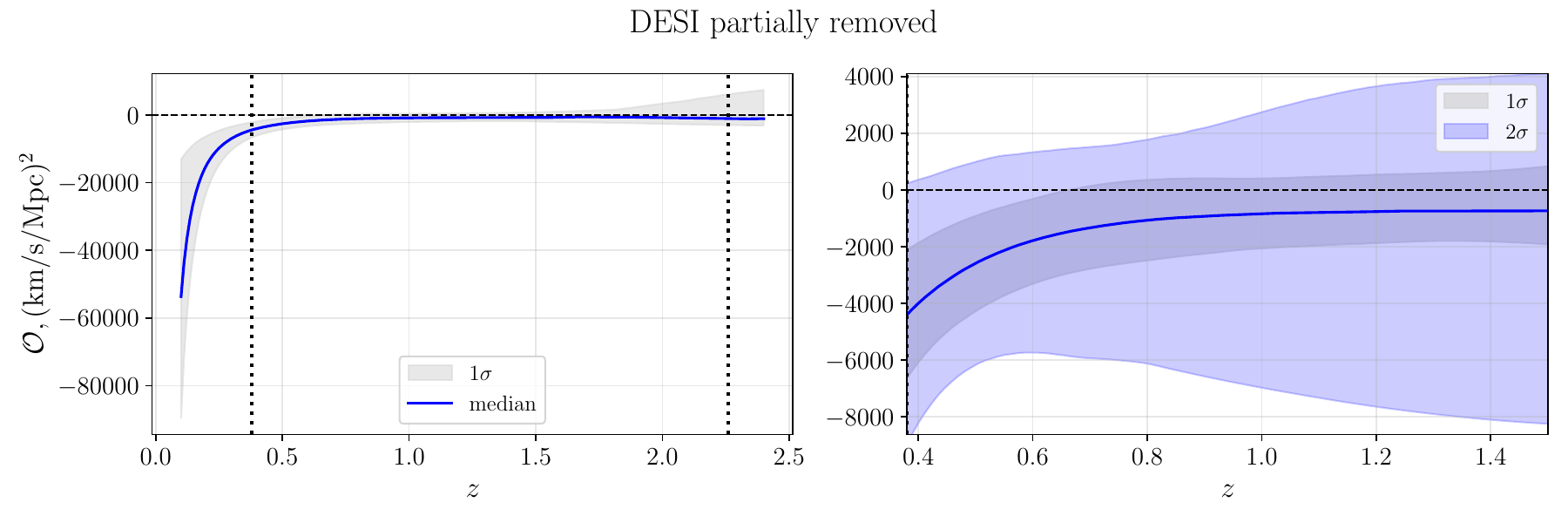}
    \caption{The $\mathcal{O}$ test (eq. \ref{eq:O}) based on supernovae and BAO data, where three DESI data points have been removed due to redshift overlap with DESI data points. We show the median and 16th/84th percentiles (labeled $1\sigma$) obtained with the 200 identified symbolic expressions combined into 40,000 $\mathcal{H}$ and $d_A$ pairs. Dashed lines are added to show the redshift interval where both datasets contain data. The right-hand plot shows a close-up and an extra uncertainty band (labeled $2\sigma$) corresponding to the 2.275/97.725 percentiles.}
    \label{fig:BOSS_O}
\end{figure*}

\begin{figure*}
    \centering
    \includegraphics[width=2\columnwidth]{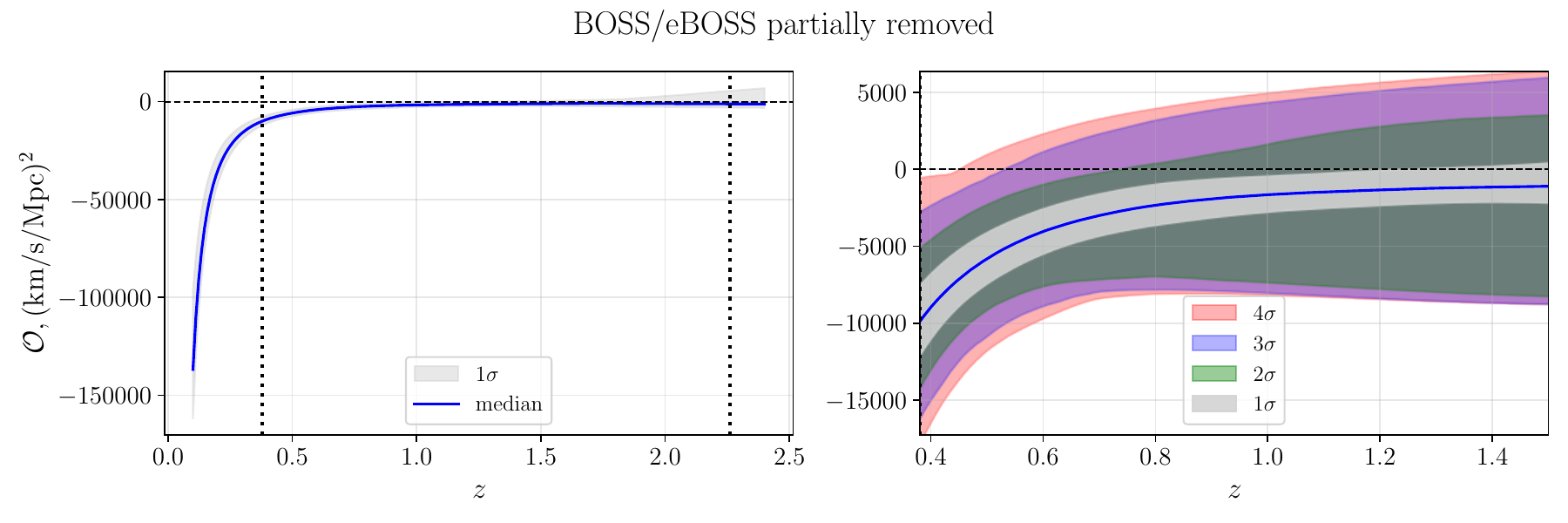}
    \caption{The $\mathcal{O}$ (eq. \ref{eq:O}) based on supernovae and BAO data, where three (e)BOSS data points have been removed due to redshift overlap with DESI data points. We show the median together with 16th/84th percentiles (labeled $1\sigma$) obtained with the 200 identified symbolic expressions. Dashed lines are added to show the redshift interval where both datasets contain data. The right-hand plot shows a close-up which includes extra uncertainty bands: a 2.275/97.725 percentiles range corresponding to $2\sigma$ for Gaussian distributed values, a 1.35th/99.86th range corresponding to $3\sigma$, and 0.135/99.87 corresponding to $4\sigma$.}
    \label{fig:DESI_O}
\end{figure*}

\noindent
In Fig. \ref{fig:BOSS_O} and \ref{fig:DESI_O} we show the constraints on $\mathcal{O}$. We see the same pattern as for the CBL-test: For the data combination where DESI data points are partially removed, the $2\sigma$ uncertainty interval agrees with the flat FLRW prediction of 0 in the entire redshift interval covered by the considered data. For the data combination where eBOSS/BOSS data is partially removed, the uncertainty  must exceed $4\sigma$ to include 0 at the lowest redshifts in the interval of the data.

\section{Diagnostic consistency relations: Numerical constraints with real data II}\label{ref:resultsII}
\begin{figure}
    \centering
    \includegraphics[width=\columnwidth]{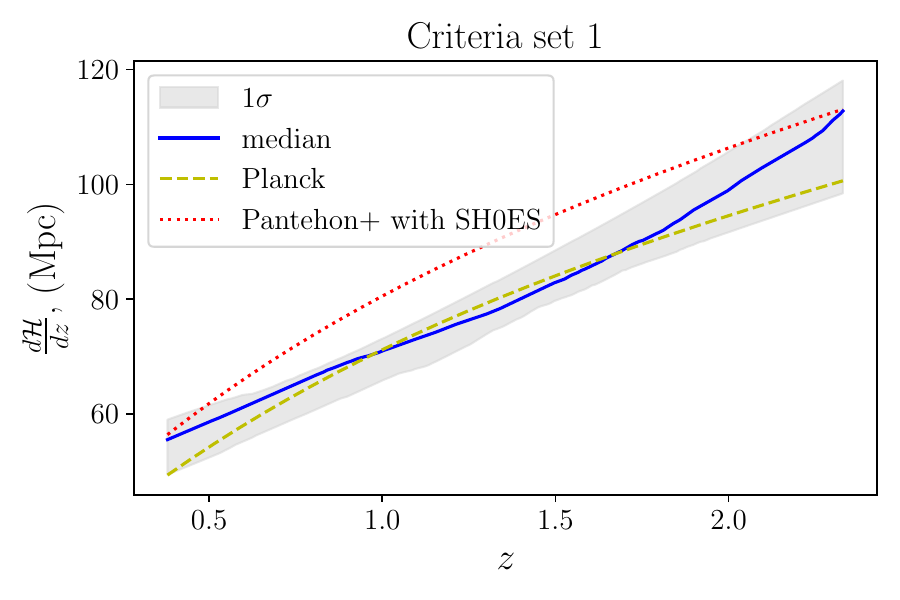}\\
    \includegraphics[width=\columnwidth]{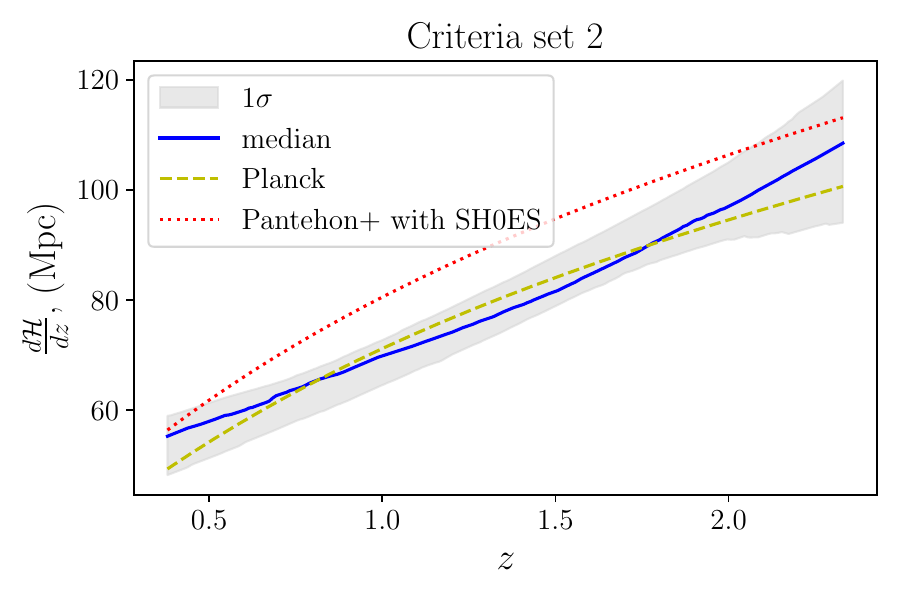}\\
        \includegraphics[width=\columnwidth]{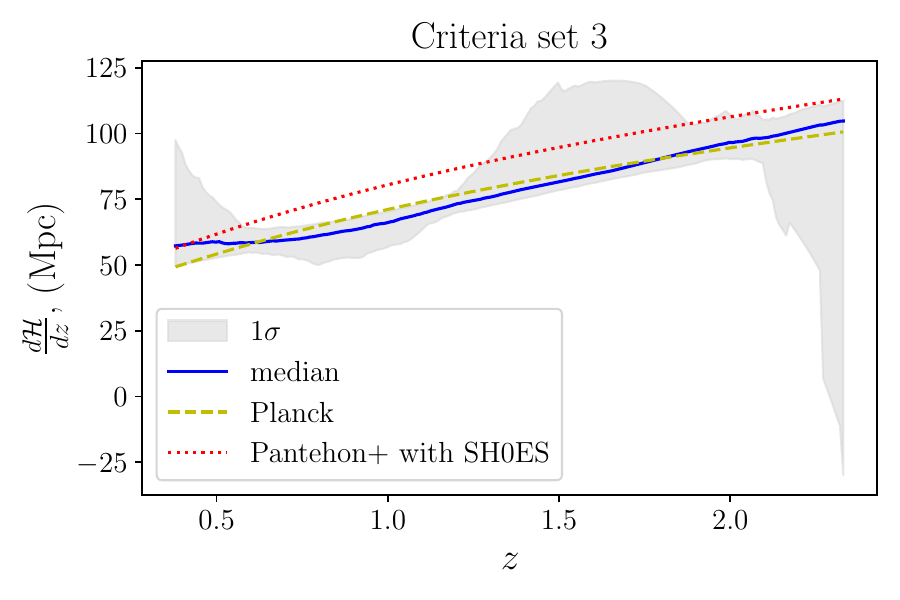}
    \caption{Reconstructed $\mathcal{H}'$ using three different sets of criteria for rejecting/retaining functions. The medians are shown together with 16th/84th percentiles and $\Lambda$CDM predictions. The results are shown together with a single bootstrap data sample. Note that spikes in the uncertainty band appear for criteria set 3 because it contains expressions that diverge at certain redshifts. 
    }
    \label{fig:Hp_criteria}
\end{figure}

\begin{figure*}
    \centering
    \includegraphics[width=2\columnwidth]{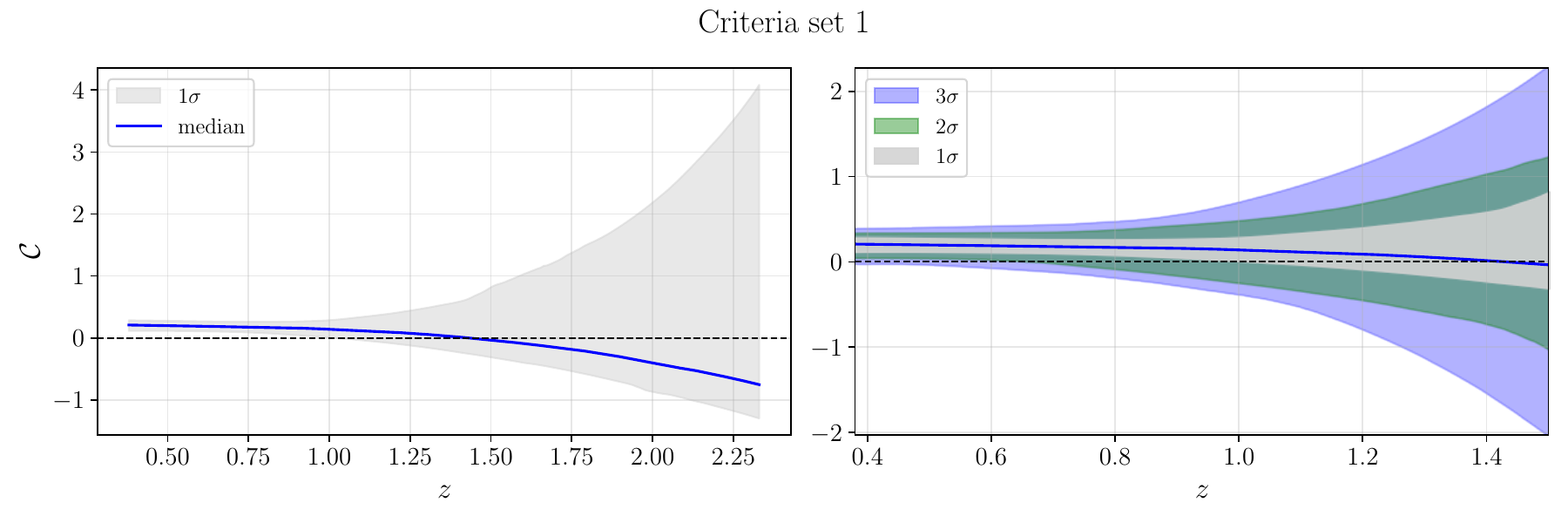}\\
    \includegraphics[width=2\columnwidth]{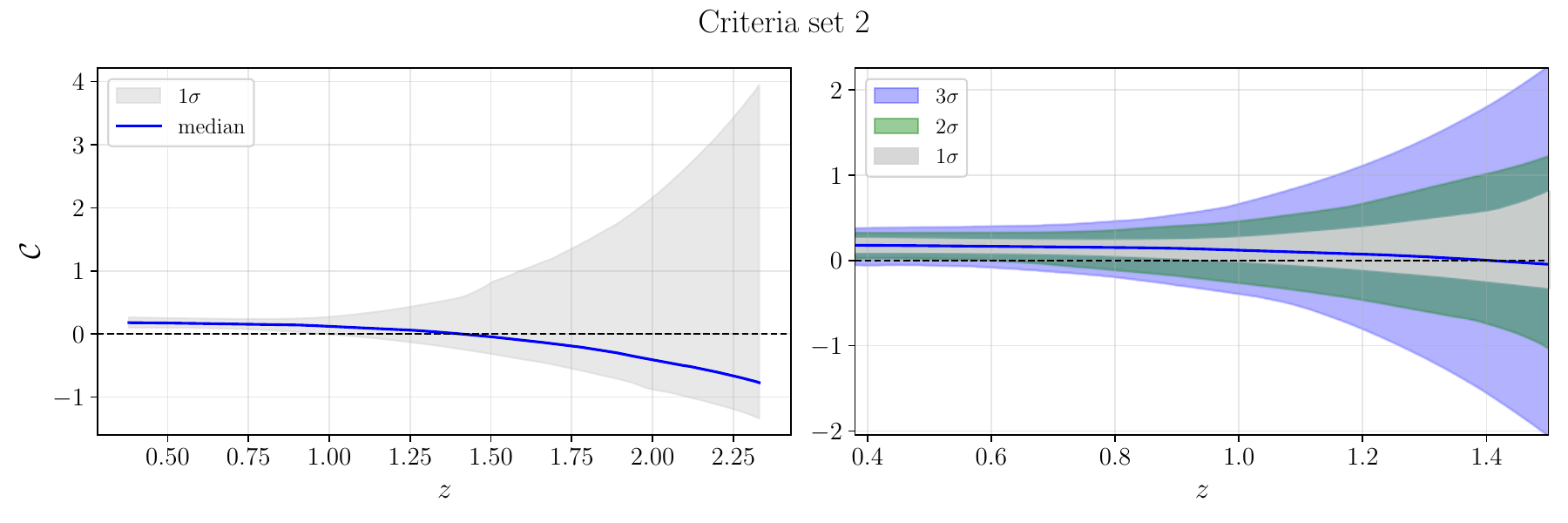}\\
        \includegraphics[width=2\columnwidth]{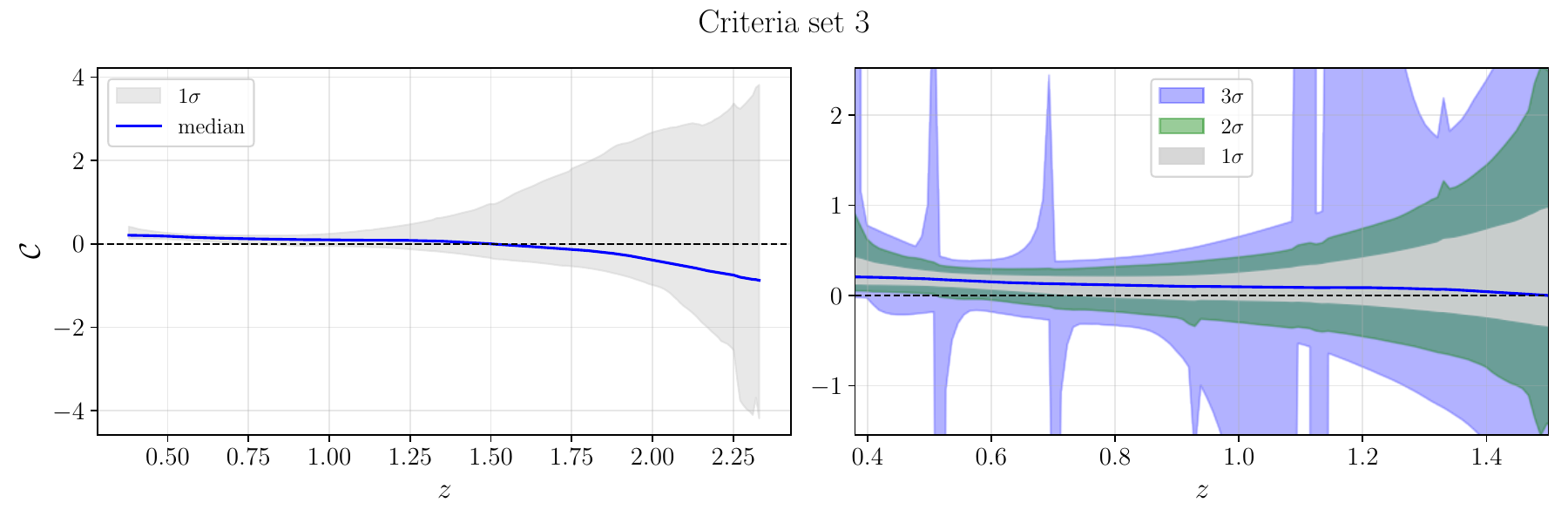}\\
    \caption{The density test $\mathcal{C}$ obtained using three different selection criteria sets. The best-fit Planck value and best-fit value based on Pantheon+ and SH0ES are shown together with the median and percentiles corresponding to $1-3\sigma$ obtained with the symbolic expressions. Results are only shown in the redshift interval that each relevant data set covers. The right-hand plot shows a close-up. Note that spikes in the uncertainty bands appear for criteria set 3 because they contain expressions that diverge at certain redshifts.  
    }
    \label{fig:C_criteri}
\end{figure*}

\begin{figure*}
    \centering
    \includegraphics[width=2\columnwidth]{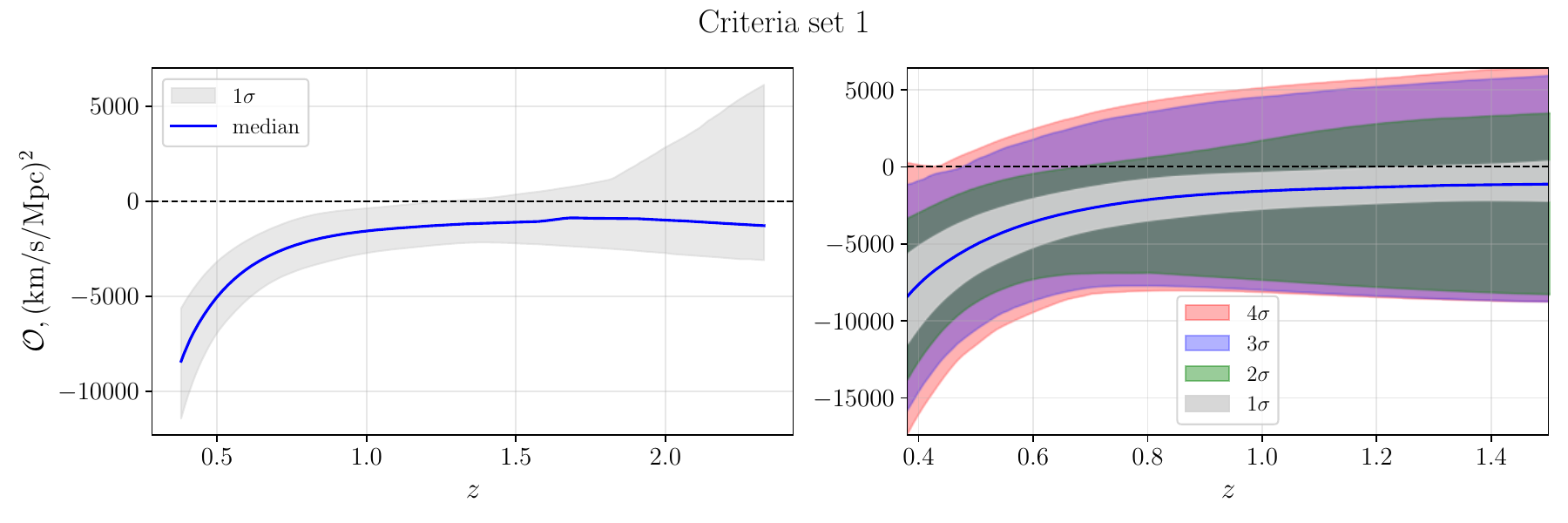}\\
    \includegraphics[width=2\columnwidth]{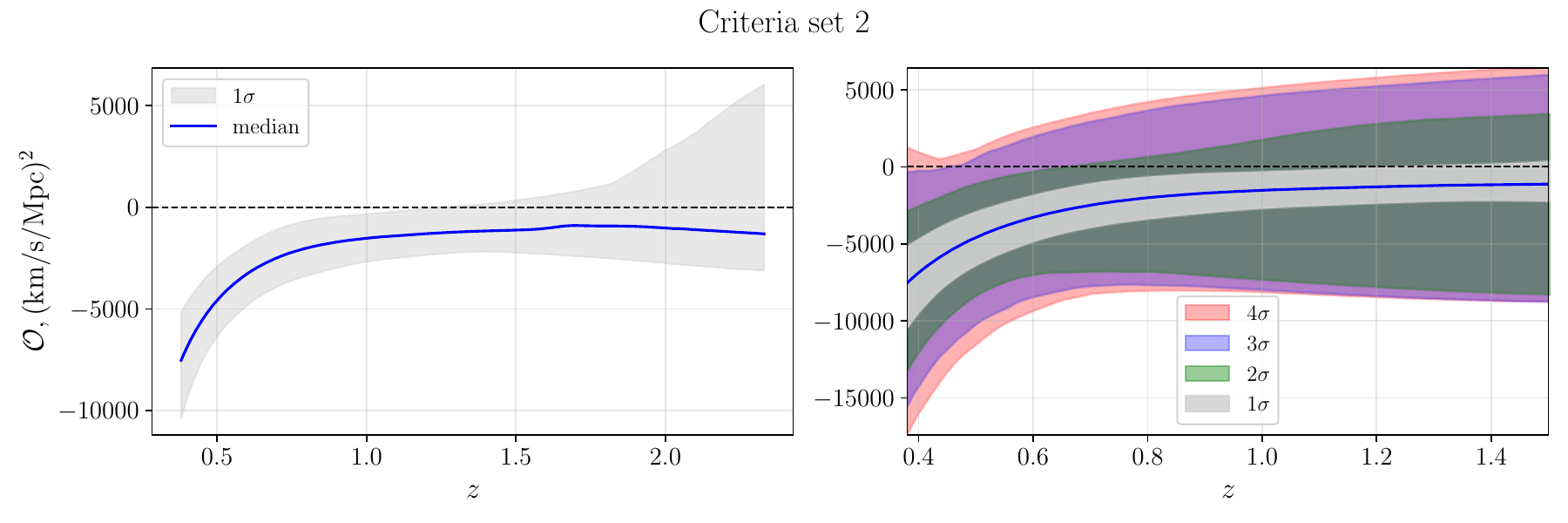}\\
        \includegraphics[width=2\columnwidth]{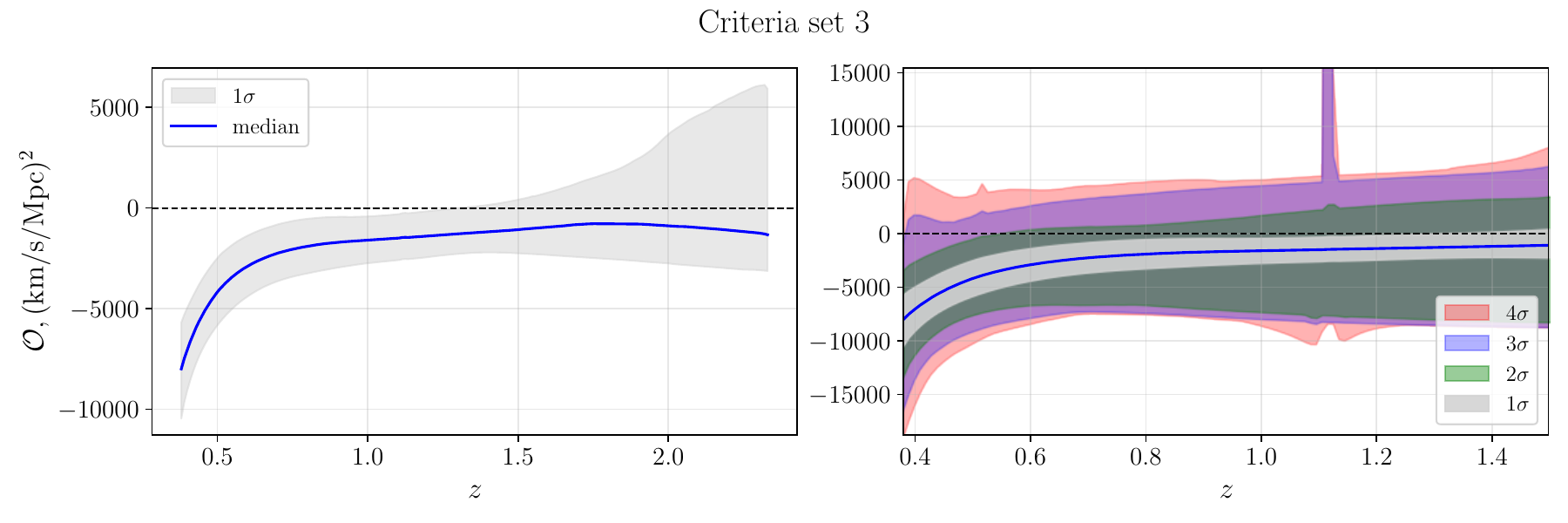}\\
    \caption{The density test $\mathcal{O}$ obtained using three different selection criteria sets. The best-fit Planck value and best-fit value based on Pantheon+ and SH0ES are shown together with the median and percentiles corresponding to $1-4\sigma$ obtained with the symbolic expressions. Results are only shown in the redshift interval that each relevant data set covers. The right-hand plot shows a close-up. Note that spikes in the uncertainty band appear for criteria set 3 because it contains expressions that diverge at certain redshifts. }
    \label{fig:O_criteri}
\end{figure*}

We now apply bootstrap-based symbolic regression to DESI data release 2 (table IV in \cite{DR2}) as well as the $z = 0.38$ redshift data point from Tab. \ref{table:BAO} and combine it with the reconstructed $d_A, d_A'$ and $d_A''$ from the previous section (i.e. we are only re-reconstructing $\mathcal{H}$ and $\mathcal{H}'$ in this section). We include the BOSS data point both to have 7 data points as in the previous analysis, and because adding this data point permits us to reconstruct $\mathcal{H}$ in a larger redshift interval. We will reconstruct $\mathcal{H}$ and $\mathcal{H}'$ with this data set using three sets of criteria in order to assess how much our results depend on the specific selection criteria. The three selection criteria we will use are:
\\\\
{\underline{\bf Criteria set 1:}} Choose the three simplest (highest Loss) expressions from each PySR pareto front/hall-of-fame, excluding only expressions that are linear or constant. Expressions must be used even if they are clearly pathological.\\
{\underline{\bf Criteria set 2:}} Choose the up to three lowest loss (highest complexity) expressions from each PySR hall-of-fame that is not pathological. We will here use ``pathological'' to mean the same clearly incorrect expressions we removed using the criteria listed in section \ref{sec:method}, i.e. expressions that are oscillating, U-shaped, divergence or not defined in parts of the training region (or where their first derivative fulfills one or more of these criteria). If three functions in a given hall-of-fame fulfill the criteria, all three functions must be used. Otherwise, the number of functions (0-2) not being pathological must be selected.\\
{\underline{\bf Criteria set 3:}} For each hall-of-fame, choose the single expression minimizing the quantity $\rm Loss + Penalty\times Complexity$, where Loss and Complexity are provided and defined by PySR, and the Penalty can be used to vary the relative weight of Loss and complexity of the expressions. We choose a penalty of 1.0 so that Loss and complexity are weighted equally.
\\\\
In practice, selection criteria set 2 is nearly identical to the selection criteria listed in section \ref{sec:method}. The main difference is that the former criteria of putting a threshold on the Loss has now been replaced by requiring rejection of constant and linear functions. For criteria set 1 we note that in practice, the requirement of including even pathological functions was not necessary as all the simplest functions in the halls-of-fame turned out to be non-pathological. On the surface, selection criteria set 3 looks very different from the two other criteria sets since it is based on a mathematical weighting of function complexity and accuracy which makes it possible to automate (as we have done) while the other two criteria sets must be carried out through manual inspection of halls-of-fame. 
Although criteria set 3 may look more objective than the others, we remind the reader that it requires several subjective inputs, namely the choice of measures of accuracy and complexity as well as the choice of penalty size. Here we choose to use the Loss and complexity measures provided directly by PySR and the ``default'' penalty of 1. Other choices would lead to other final families of functions.
\newline\indent
While it is easy to implement a mathematical  $\rm Loss + Penalty\times Complexity$ criteria for PySR output, this is not the case for the output from cp3-bench where multiple algorithms are used, each providing results in different formats. Although cp3-bench provides a uniform mse measure for the ``best fit'' expressions selected by each algorithm, it does not provide a uniform measure of complexity (and the measures provided by each algorithm are not all the same).
\newline\newline
Fig. \ref{fig:Hp_criteria} shows the reconstructed $\mathcal{H}'$. The two reconstructions from criteria sets 1 and 2 are very similar, but the uncertainty band is visibly broader at high redshifts when using criteria set 2. Selection criteria set 3 clearly leads to a different uncertainty band. The difference is not as big as one might expect based on the very different family of functions obtained by this set of selection criteria which e.g. led to 2.5 \% of the selected expressions to be pathological, being undefined in parts of the considered redshift region. The impact of these expressions is most easily seen in the appearance of spikes in the uncertainty bands. We furthermore see that even the ``straight''/non-oscillatory part of the uncertainty band obtained with criteria set 3 is shifted a small amount compared to the uncertainty bands of two other criteria set. This emphasizes that different selection criteria affects the resulting constraints as we also see for the constraints discussed below.
\newline\indent
Since the reconstructed $\mathcal{H}$ are very similar, we do not show them here. We similarly find only small differences in the reconstructed $\mathcal{M}$ for which the results are all very similar to those reported in the previous section and which we therefore omit here. Instead, we move on to show the resulting constraints on $\mathcal{C}$ in Fig. \ref{fig:C_criteri}. In this case, a scrutinization of the close-ups show that the uncertainty bands are slightly broader when using criteria set 2 than criteria set 1, and we see a clear difference for selection criteria 3, where the FLRW expectation lies within the $2\sigma$ uncertainty band for larger redshift intervals than when using the two other, more subjective, criteria.
\newline\indent
Lastly, in Fig. \ref{fig:O_criteri}, we show the constraints on $\mathcal{O}$. Here we again find that the uncertainty bands are larger when using criteria set 2 compared to 1. In this particular case, the difference between the two is more important because the FLRW violation is nearly brought down to within $3\sigma$ in the entire relevant redshift interval when using criteria set 2. For criteria set 3, the bands have widened to the extent that the flat FLRW expectation is within $3\sigma$ everywhere in the considered redshift interval.
\newline\newline
The results presented in this section demonstrate that using bootstrap-based symbolic regression to asses uncertainties is qualitatively robust while the quantitative uncertainties are sensitive to the exact selection criteria. The latter is especially important in cases as in Fig. \ref{fig:C_criteri} and \ref{fig:O_criteri}, where it seems that only smaller changes in the variance of the selected functions is required to shift the FLRW violation from $3-4\sigma$ to $2\sigma$. Despite the quantitative shifts when using different criteria set, we emphasize that the results are remarkably robust considering the differences in the three criteria sets. Most importantly, we repeat that criteria set 3 lead to 2.5 \% of the final functions to be non-defined on parts of the considered redshift interval and divergent close to these regions. These functions were all removed by our ``common sense'', manual inspections in criteria set 1 and 2.
\newline\newline
For all three diagnostic consistency relations, the constraints obtained in this section are very similar to those obtained by the DESI-dominated results of section \ref{sec:resultsI}. This is not too surprising since we are also here using DESI dominated data. Most importantly, however, we note that the FLRW violation is less severe when using the newer DESI data.

\section{Conclusion}\label{sec:conclusion} 
In this work, we introduced bootstrap-based symbolic regression as a method to obtain model-independent reconstructions of $d_A, d_A', d_A'', \mathcal{H}$ and $\mathcal{H}'$ with uncertainty. For demonstration and to study the robustness of the method, we applied it to supernova and BAO data. We first considered mock observations based on FLRW models to demonstrate the ability of the method to correctly reconstruct $d_A, d_A', d_A'', \mathcal{H}$ and $\mathcal{H}'$ as well as our newly introduced diagnostic FLRW consistency relations within one standard deviation. We then moved on to explore the equivalent constraints using real data from BOSS, eBOSS, DESI and Pantheon+ with SH0ES. Our newly introduced density-test, $\mathcal{M}$, is not well-constrained with current amount of data, and both Planck and Pantheon+ with SH0ES $\Lambda$CDM predictions are within one standard deviation of our median result. When constraining our generalized Clarkson-Basses-Lu test, $\mathcal{C}$ we find disagreement with the FLRW prediction $\mathcal{C} = 0$ in large parts of the redshift interval examined, and for all data combinations and selection criteria. The disagreement is most significant for data combinations for which the BAO data is dominated by DESI data release 1 measurements, where the disagreement exceeds $3\sigma$ up to $z\approx 0.5$. When considering DESI data release 2, the FLRW prediction is within $2\sigma$ of the median result for the main part of the considered redshift interval. 
\newline\indent
We finally considered the integrated version of the test statistic $\mathcal{C}$, namely $\mathcal{O}$, where we also find significant deviation from flat FLRW expectations and with a slope of $\mathcal{O}$ in $z$ that is indeed compatible with the original constraints on $\mathcal{C}$. When considering DESI data release 1, the violation reaches just above $4\sigma$ at the lowest part of the relevant redshift interval. When considering DESI data release 2, the result shows a $3-4\sigma$ deviation from the flat FLRW expectation. 
The results for $\mathcal{O}$ (and for $\mathcal{C}$) are well approximated by the form $\mathcal{O} \approx c_1/D^2$ in all test cases for the redshift range where we have data coverage, where $c_1$ is a negative constant. From the formula for $\mathcal{O}$, this implies that $(\mathcal{H} D')^2 \approx 1 + c_1$. We may think of $c_1$ as a calibration offset of $D$ relative to the radial distance $1/\mathcal{H}$. For $c_1= - 0.2$, which appears to lie within the $2\sigma$ bound of all data combinations and selection criteria, this corresponds to a calibration offset of $D$ relative to $1/\mathcal{H}$ of $-10\%$, which is exactly the calibration offset between Pantheon+ and the BAO data (calibrated with the CMB). Thus, our results are quantitatively meaningful in light of the Hubble tension.
\newline\indent
In order to examine the robustness of our results against subjective selection criteria, we consider DESI data release 2 with three different sets of selection criteria in our bootstrap-based symbolic regression method. 
A comparison between the three sets of resulting constraints reveals that although our method is qualitatively robust, the quantitative uncertainties do depend on the selection criteria at a level that is important for precisely quantifying the significance of the FLRW violation. Nonetheless, the consistent violation of the flat FLRW consistency relation at $2-4\sigma$ that we do see across datasets and selection criteria is intriguing as it may hint at a non-FLRW solution to the Hubble tension (see \cite{Bolejko:2017fos, Heinesen:2020sre,Clifton:2024mdy} for simple examples of such solutions) and/or the apparent dynamical-dark-energy findings driven by DESI (see \cite{Camarena:2025upt,Ginat:2026fpo}).
\newline\indent
Across selection criteria and data sets, we obtain at least a $2\sigma$ violation of FLRW expectations. In this sense, our results are to the best of our knowledge the first significant detections of violation of FLRW self-consistency. However, our results are based on a newly introduced method which we denote bootstrap-based symbolic regression. Although our results suggest that the method is qualitatively robust, it depends on subjective choices in terms of hyperparameters, binary operators and functions. Further work toward refining, and building community consensus on, appropriate uncertainty estimation in symbolic regression is required.
\newline\indent
If the violations we have found  here are robust, it ultimately means that most cosmological models considered as candidates for solving the tensions of the $\Lambda$CDM model (e.g., dynamical/interacting/new dark energy models, modified gravity, and running of natural constants within the FLRW framework) are ruled out. This would alter the way that cosmologists should approach the construction of solutions to cosmological tensions.
Alternatively, if one believes that a cosmological solution must necessarily fall within the FLRW class of spacetimes, the only viable explanation for the tensions appear to be of astrophysical character or exotic early-time mechanisms that invalidate current standard BAO interpretation. Irrespective of the interpretation of our findings, further examinations are required to establish their robustness.
The methods in this paper could be applied to other FLRW/$\Lambda$CDM consistency tests proposed in the literature \cite{Rasanen:2014mca,list}, which could yield complementary valuable information.
\newline\indent
Data is not yet precise and ample enough to secure tight constraints on our model-independent density constraint that we presented in the companion letter \cite{PRL}. Especially more high-precision BAO data (or better constrained cosmic chronometer measurements) is missing in order to obtain sky-averaged estimates with modest uncertainty bands. However, this is expected to be achieved over the next years, and our results demonstrate that our approach enables a direct test of the $\Lambda$CDM model without introducing cosmological parameter fitting: any deviations from $\Lambda$CDM predictions can be directly interpreted in terms of a physical quantity (the density field) rather than simply as an ``anomaly''. Ultimately, the presented framework provides a pathway towards genuinely model-independent cosmology, with tests that offer clear physical insight into the origin of tensions with $\Lambda$CDM predictions.
\newline\indent
Although we have focused on the sky-averaged density field as a function of the redshift, the method can also be used to map density fluctuations across the sky once data becomes sufficiently ample. More data will also make it possible to constrain other interesting quantities such as the optical expansion (measuring the isotropic image distortion in the direction from source to observer and setting $E_0 = 1$) given by $\hat\theta = -2\frac{d_A'}{d_A}(1+z)^2\mathcal{H}$ which could complement standard lensing maps.
\newline\newline
A final remark concerns our treatment of uncertainties through bootstrap sampling. Because symbolic regression typically does not natively provide uncertainty estimates, some form of data resampling appears indispensable for most symbolic regression algorithms. Bootstrapping perturbs the data within their measurement uncertainties, compels the regression procedure to explore multiple plausible functional forms, and naturally reveals which reconstructed features are robust and which arise from noise. The main limitation of our current bootstrap-based approach lies in the subjectivity of the selection criteria used to reject expressions. Although we have shown that reasonable variations in these criteria have only a small impact on the final results, an ideal framework would replace such choices with more objective schemes. We take a small step in this direction with our selection criteria set 3 in section \ref{ref:resultsII}. Even though this set of criteria is fully automated, it is not fully objective as it still requires the user to choose weights and measures of accuracy and complexity. Furthermore, it represents a heuristic combination of Loss and accuracy. A possible direction of improvement could be to introduce a selection criteria similar to the ranking scheme described in \cite{SR5}. Indeed, using the ranking scheme of \cite{SR5} could be used to generate uncertainties even without bootstrapping, which could be an interesting path forwards to further refine the method of combining symbolic regression with error estimates presented here.

\begin{acknowledgments}
The authors thank Harry Desmond and Pedro G. Ferreira for correspondence which inspired the introduction of criteria set 3, and Chris Clarkson for comments on our draft. SMK acknowledges funding by Villum Fonden, grant VIL53032. 
AH is supported by the Perren Fund at the University of London, and wishes to thank the Astronomy Unit at Queen Mary University of London for their support.
Part of the numerical work was performed using the UCloud interactive HPC system managed by the eScience Centre at the University of Southern Denmark. Our numerical codes were written in Python and utilized Numpy\footnote{https://numpy.org/} \cite{Harris_2020}, Matplotlib\footnote{https://matplotlib.org/} \cite{matplotlib}, Pandas\footnote{https://pandas.pydata.org/}, sympy\footnote{https://www.sympy.org/en/index.html}, pathlib\footnote{https://docs.python.org/3/library/pathlib.html}, json\footnote{https://docs.python.org/3/library/json.html}, argparse\footnote{https://docs.python.org/3/library/argparse.html}, subprocesses\footnote{https://docs.python.org/3/library/subprocess.html}, sys\footnote{https://docs.python.org/3/library/sys.html}, os\footnote{https://docs.python.org/3/library/os.html} and io\footnote{https://docs.python.org/3/library/io.html}. Debugging and data handling was assisted by ChatGPT.
\newline
{\bf Author contribution statement}: Analytical derivations were performed by SMK and independently verified by AH. The numerical work was conducted by SMK. SMK led the overall development of the project, with both authors making substantial contributions to the refinement of the work and interpretation of results. Both authors contributed significantly to the writing of the manuscript.
\end{acknowledgments}

\appendix
\section{Hyperparameters}\label{app:pysr}
In this appendix, we provide the details of the hyperparameter settings chosen for AIFeynman, QLattice, ITEA, and GeneticEngine when constraining $d_A$, as well as the choices of hyperparameter we implemented for PySR when constraining $\mathcal{H}$.
\newline\newline
Note that the algorithms work in fundamentally different ways and therefore have different hyperparameters that can be set. When setting hyperparameters for AIFeynman, QLattice, ITEA and GeneticEngine we modify the \texttt{procedure.py}-files in cp3-bench. For AIFeynman we use the following hyperparameter settings:
\begin{itemize}
    \item The try time is limited to 8000.
    \item The maximum polynomial degree is set to 5.
    \item The number of neural network epochs is set to $10^5$.
    \item ``maxtime'' is set to 12000.
\end{itemize}
For QLattice we set the number of epochs to 8000 and the complexity maximum to 8.
\newline\indent
For ITEA we use the following hyperparameter settings:
\begin{itemize}
    \item Population number is set to 1000.
    \item Number of generations is set to 7000.
    \item ``exponents'' is limited to the interval -5 to 5.
    \item ``termlimit'' is given the interval 1 to 5.
    \item We permit up to 20 non-zero exponents.
    \item ``transfunctions'' is given the options ``Id'', ``Sin'', ``Cos'', ``SqrtAbs'', ``Log'' and ``Exp''.
\end{itemize}
For GeneticEngine we use the following hyperparameter settings:
\begin{itemize}
    \item Population size is set to 500.
    \item ``n\_elites'' is set to 20.
    \item ``n\_novelties'' is set to 40.
    \item Number of generations is limited to 2000.
    \item The maximum depth is set to 14.
    \item We set ``favor\_less\_deep\_trees'', ``hill\_climbing'' and ``Time\_stop\_criteria'' to True.
    \item We set the probability for mutation to 0.2.
    \item We set the time limit to 14400.
\end{itemize}
When using PYSR to constrain $\mathcal{H}$, we use it outside of cp4-bench. Thus, we set the hyperparameters through PySR's \texttt{PySRregressor} function. We use the following hyperparameters:
\begin{itemize}
    \item Population size is set to 90 (larger population size leads to greater diversity).
    \item Number of cycles per iteration is set to 400
    \item Number of iterations is set to 800
    \item We set the algorithm to stop early if a function fulfilling the condition that the loss is below $10^{-6}$ and its complexity is less than $10$ is found.
    \item We set the maximum runtime to be one hour
    \item We set maxsize (limits the complexity) to be 20 and maxdepth (limits nesting of functions) to be 10
    \item We permit the binary operators $\cdot, +, -$
    \item We permit the unary operators ``square'', ``cube'', ``exp'' and ``log'' and limit the complexity of the arguments in each of these to be 9.
    \item Lastly, we limit the nestedness of the operators to 1 except for exponentials inside exponentials which we do not permit, and logarithms inside logarithms which we also do not permit.
\end{itemize}
The above choices are largely small modifications of default parameters which we adjusted only mildly in order to minimize the fraction of returned symbolic expressions that are undefined in parts of the considered redshift interval and to reduce computation time. For PySR we explicitly removed division from the binary operators in order to reduce the fraction of symbolic expressions that contained division by zero within the redshift interval of our data.

\bibliography{bibliography}

\end{document}